\documentclass[twocolumn,epjc3]{svjour3}
\journalname{Eur. Phys. J. C}
\usepackage{mathrsfs,amsmath,latexsym,amssymb,amsfonts,enumerate,graphicx,txfonts,subfigure,graphics,cite}
\usepackage{xcolor}
\usepackage[colorlinks,linkcolor=blue,citecolor=blue,urlcolor=blue,plainpages=false,pdfstartview=FitH]{hyperref}
\usepackage{etoolbox,orcidlink}
\allowdisplaybreaks
\newcommand{\GZU}{School of Physics, Guizhou University, Guiyang 550025, China}
\usepackage{multirow,booktabs}
\usepackage{bm}

\begin{document}

\title{Gravitational waveforms and accretion characteristics in a quantum-corrected black hole without Cauchy horizons}

\author{Shilong Huang
	\thanksref{addr1,e1}
	\orcidlink{0009-0002-5996-2952}
	\and Jiawei Chen
	\thanksref{addr1,e2}
	\orcidlink{0009-0003-5390-8186}
	\and Jinsong Yang
	\thanksref{addr1,e3}
	\orcidlink{0000-0003-4051-2767}
}

\institute{\GZU\label{addr1}}

\thankstext{e1}{e-mail: gs.slhuang24@gzu.edu.cn}
\thankstext{e2}{e-mail: gs.chenjw23@gzu.edu.cn}
\thankstext{e3}{e-mail: jsyang@gzu.edu.cn (corresponding author)}

\date{Received: date / Accepted: date}
\maketitle

\begin{abstract}
The use of physical phenomena in the strong-field regime has become a primarily methodology for probing quantum-corrected gravity. This paper investigates periodic orbits, gravitational waves, and accretion disk radiation for a quantum-corrected black hole without Cauchy horizons. First, by analyzing the trajectory equations of massive particles in the equatorial plane, we study the influence of the quantum parameter $\zeta$ on the stability of circular orbits. The results show that an increase in $\zeta$ leads to an outward migration of both the innermost stable circular orbit and the marginally bound orbit, accompanied by an increase in the required specific angular momentum for particle motion on these two orbits. Then, we further investigate the periodic orbit characteristics of particles and compute the associated gravitational waveforms for extreme mass-ratio inspirals. It is demonstrated that quantum corrections induce a cumulative phase shift in the gravitational wave signal, leading to significant dephasing compared to the classical Schwarzschild case. Furthermore, based on the Novikov-Thorne thin accretion disk model, we evaluate the radiation characteristics of the accretion disk around this quantum-corrected black hole. The results indicate that the introduction of the quantum parameter suppresses the radiant energy flux, effective temperature, and overall radiative efficiency of the disk. These distinctive dynamical and radiative deviations provide potential phenomenological support for distinguishing quantum-corrected geometries from classical black holes using multiple observational means in the future.
\end{abstract}
\maketitle

\section{Introduction}

General relativity (GR) serves as the foundational framework for our current description of gravity, having passed numerous experimental tests across scales from the solar system to strong-field environments, e.g., black holes (BHs)~\cite{Einstein:1915bz,Einstein:1937qu,Peters:1964zz,Bondi:1962px}. Observations, including the detection of gravitational waves by the LIGO and Virgo collaborations~\cite{LIGOScientific:2016aoc,LIGOScientific:2016vbw,LIGOScientific:2016vlm,LIGOScientific:2016emj,LIGOScientific:2018mvr,LIGOScientific:2020aai} and the imaging of supermassive BHs via Event Horizon Telescope (EHT)~\cite{EventHorizonTelescope:2019dse,EventHorizonTelescope:2022wkp}, have further confirmed that the BHs predicted by GR are real astrophysical objects, not merely theoretical constructs. Nonetheless, classical GR faces the singularity problem, where predictions fail due to unbounded spacetime curvature~\cite{Penrose:1964wq,Hawking:1970zqf}. Together with the information loss paradox, this indicates that GR is an effective theory likely to be replaced by a more complete quantum gravity description at small scales.

To address singularities, several approaches have been developed~\cite{Hartle:1983ai,Strominger:1996sh,Bojowald:2008zzb,Ashtekar:2013hs}. Notably, loop quantum gravity (LQG) provides a non-perturbative quantization of spacetime geometry~\cite{Rovelli:1997yv,Han:2005km,Perez:2012wv,Yang:2016kia,Long:2020agv,Zhang:2022vsl}. In cosmology, LQG-inspired effective models replace the classical singularity with a quantum bounce~\cite{Ashtekar:2003hd,Bojowald:2005epg,Ashtekar:2006wn,Ding:2008tq,Yang:2009fp}. In the field of BH research, LQG has also yielded a series of results in addressing BH singularities~\cite{Ashtekar:2005qt,Gambini:2020nsf}. However, in the Hamiltonian formulation, the construction of effective BHs that are both generally covariant and consistently incorporate quantum corrections from loop quantum gravity remains a persistent theoretical challenging. Recently, by impsing diffeomorphism invariance at the effective level, researchers have obtained a new family of quantum-corrected black hole (QCBH) solutions that strictly satisfy general covariance~\cite{Zhang:2024khj,Zhang:2024ney}. Notably, the newly derived QCBH solutions lack Cauchy horizons, a property that may naturally circumvent the mass inflation instability~\cite{Zhang:2024ney}.

To connect theoretical models with astrophysical data, it is crucial to investigate how such quantum-corrected affect observations. Two of the approaches for this are studying the gravitational wave radiation generated by the motion of test-particle along periodic orbits and analyzing accretion disk radiation. Future space-based detectors such as Taiji, TianQin, and LISA will primarily target the process in which a stellar-mass compact object spirals into a BH, i.e., extreme mass-ratio inspirals (EMRI) system~\cite{Hu:2017mde,TianQin:2015yph,Gong:2021gvw,Danzmann:1997hm}. The gravitational waves emitted during this inspiral phase carry detailed imprints of the underlying spacetime geometry. Therefore, it is necessary to study the orbits in this inspiral phase, among which periodic orbits, displaying a characteristic “zoom-whirl” pattern~\cite{Levin:2008mq}, are particularly important. Subtle corrections to the spacetime due to quantum effects can be uncovered by examining deviations in the orbital frequencies and phase evolution of these paths.

In addition to the direct detection means provided by gravitational-wave probes, the emission from accretion disks offers a direct observational probe into the spacetime near the BH. According to the standard geometrically thin, optically thick disk model developed by Novikov and Thorne, the radiation characteristics of the accretion disk depend on physical quantities such as the specific angular momentum, specific energy, and the innermost stable circular orbit (ISCO) of particles~\cite{Page:1974he,Thorne:1974ve}. Therefore, any change in spacetime geometry that could alter these physical quantities will necessarily leave observable signatures in the accretion flow’s spectral energy distribution (SED), luminosity, and radiative efficiency.

In this work, we perform a systematic analysis of massive particle orbits and accretion processes around the aforementioned diffeomorphism-preserving QCBH, which lacks Cauchy horizons. Our goal is to quantify how the quantum parameter $\zeta$ alters the effective potential for massive particles, thereby changing the stability limits for circular orbits and modifying the properties of periodic orbits. Using a numerical kludge scheme~\cite{Babak:2006uv}, we generate gravitational waveforms to evaluate how well this quantum-corrected model can be distinguished from the standard Schwarzschild case. We also compute the energy flux and temperature profiles of thin accretion disks to determine whether quantum effects increase or decrease the radiative output.

The paper is structured as follows. Section~\ref{section2} introduces the QCBH metric and derives the effective potential and orbital equations, laying the foundation for the analysis of bound and periodic orbits. Section~\ref{sec:periodic} studies the general characteristics of bound, circular, and periodic orbits, providing the basis for gravitational wave and accretion disk analysis. Section~\ref{sec:gw} presents the computed gravitational waveforms from extreme mass-ratio inspirals. Section~\ref{accretion} examines the electromagnetic signatures of thin accretion disks, with particular attention to flux distributions and energy conversion efficiency. Finally, Sect.~\ref{conclusion} summarizes our key results and discusses their implications for future observational tests. Unless otherwise specified, we set $G=c=M=1$.

\section{Quantum-corrected BH spacetime and geodesic motion of massive particles}\label{section2}

\subsection{Quantum-corrected BH without Cauchy horizons}

A common challenge in the Hamiltonian formulation of gravity theories is maintaining general covariance. Addressing this, an effective Hamiltonian constraint $H_{\rm eff}$ that preserves diffeomorphism invariance was formulated in~\cite{Zhang:2024khj,Zhang:2024ney}, independent of any gauge choice. By fixing the free functions to encode quantum gravity corrections, this framework leads to a novel class of solutions. These describe spacetimes that manifest as either regular BHs or traversable wormholes, contingent on the parameter values~\cite{Zhang:2024ney}. A key distinction from prior QCBH models is the absence of Cauchy horizons in these solutions, a feature that may bolster their stability under perturbations. The corresponding line element for the QCBH is given by~\cite{Zhang:2024ney}:
\begin{equation}
	{\rm d}s^{2} =-g_{tt}{{\rm d}t}^{2} + g_{rr}{{\rm d}r}^{2} + g_{\theta\theta}{{\rm d}\theta}^{2} + g_{\varphi\varphi}{{\rm d}\varphi}^{2},\label{line_element}
\end{equation}
where
\begin{equation}
 \begin{split}
 g_{tt} &= 1 - (-1)^{n}\frac{r^{2}}{\zeta^{2}}\arcsin\left(\frac{2M\zeta^{2}}{r^{3}}\right) - \frac{n\pi r^{2}}{\zeta^{2}},\\
 g_{rr}&= \frac{1}{g_{tt} {\left(1-\frac{4M^2\zeta^4}{r^6}\right)}},\\
 g_{\theta \theta}&=r^2, \\
 g_{\varphi \varphi}&=r^2 \sin^2{\theta}.
 \end{split}\label{metric_function}
\end{equation}
In this context, $\zeta$ denotes the quantum parameter, while $M$ represents the BH mass, and $n$ is an arbitrary integer.

\begin{figure}[htbp]
	\centering
	\includegraphics[width=8cm]{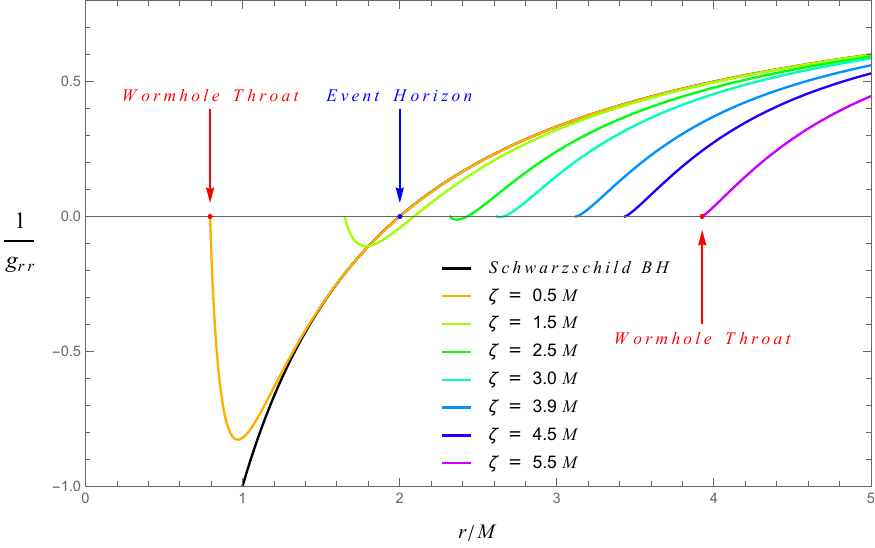}

	\caption{The location and relationship between the event horizon and the wormhole throat ($r_{\rm m} = (2M \zeta^2)^{1/3}$) in the quantum-corrected spacetime for specific values of $\zeta$. The vertical axis displays $1/g_{rr}$.}
	\label{fig:grr_inverse}
\end{figure}

Our analysis focuses on the specific choice $n=0$, associated with asymptotically flat geometries. For $\zeta <\sqrt{\frac{\pi^3}{2}}~M\cong3.937~M$, the metric corresponds to a regular BH, in which the typically singular region is supplanted by a wormhole whose throat is situated at $r_{\rm m} = (2M \zeta^2)^{1/3}$. As shown in Fig.~\ref{fig:grr_inverse}, as $\zeta$ increases beyond this bound, the QCBH horizon dissolves, and the spacetime morphs into that of a traversable wormhole. Therefore, in the subsequent study of this paper, the selection of the quantum parameter is restricted to $\zeta < 3.937~M$.

\subsection{Radial motion of massive particles in the equatorial plane}

Metric \eqref{line_element} describes a spherically symmetric spacetime endowed with the timelike and spacelike Killing vectors $(\partial / \partial t)^a$ and $(\partial / \partial \varphi)^a$, governing time translation and rotation symmetries, respectively. Considering motion in the equatorial plane ($\theta = \pi/2$), where the metric components $g_{tt}$, $g_{rr}$, $g_{\theta\theta}$, and $g_{\varphi\varphi}$ reduce to functions of the $r$. The specific energy $\tilde{E}$ and specific angular momentum $\tilde{L}$ with affine parameter $\tau$ take the form~\cite{Wald:1984rg,Liang:2023ahd}
\begin{equation}
	\begin{split}
		&\tilde{E}:=-g_{a b}\left(\frac{\partial}{\partial t}\right)^{a}\left(\frac{\partial}{\partial \tau}\right)^{b}=g_{tt}\dot{t},\\
		&\tilde{L}:=g_{a b}\left(\frac{\partial}{\partial \varphi}\right)^{a}\left(\frac{\partial}{\partial \tau}\right)^{b}=g_{\varphi \varphi}\dot{\varphi}.
	\end{split}\label{energy_and_angular_monmentum}
\end{equation}
For timelike geodesics, the four-velocity satisfies the normalization condition $g_{ab}\left( \frac{\partial}{\partial\tau} \right)^{a}\left( \frac{\partial}{\partial\tau} \right)^{b}=- 1$. Substituting the metric components and the expressions for $\dot{t}$ and $\dot{\varphi}$ derived from Eq.~\eqref{energy_and_angular_monmentum}, we obtain
\begin{equation}
		-1 = g_{tt}\dot{t}^2 + g_{rr}\dot{r}^2 + g_{\varphi\varphi}\dot{\varphi}^2 = -\frac{\tilde{E}^2}{g_{tt}} + g_{rr}\dot{r}^2 + \frac{\tilde{L}^2}{g_{\varphi\varphi}}.
\end{equation}
Solving for $\dot{r}^2$ yields the radial equation of motion for massive particles
\begin{equation}
	\dot{r}^2\equiv\left( \frac{{\rm d}r}{{\rm d}\tau} \right)^{2} = \left(1-\frac{4M^2\zeta^4}{r^6}\right)\left[\tilde{E}^{2}-V_{\rm eff}(r)\right],\label{trajectory_equation}
\end{equation}
substituting the explicit forms of $g_{tt}$, $g_{rr}$, and $g_{\varphi\varphi}$ from Eq.~\eqref{metric_function}, we obtain the final expression for the radial motion in the QCBH spacetime
\begin{equation}
 \begin{split}
		\left(\frac{{\rm d} r}{{\rm d} \varphi}\right)^2 =r^4 \left[\frac{{\tilde{E}}^2}{{\tilde{L}}^2}- \frac{V_{\rm eff}(r)}{\tilde{L}^2}\right]\left (1-\frac{4M^2\zeta^4}{r^6}\right),\label{eq:motion_equation}
 \end{split}
\end{equation}
where the effective potential is defined as
\begin{equation}
	V_{\rm eff}(r) =g_{tt}\left(1+\frac{\tilde{L}^{2}}{r^{2}}\right).\label{effective_potential}
\end{equation}

\begin{figure}[htbp]
	\centering
	\includegraphics[width=8cm]{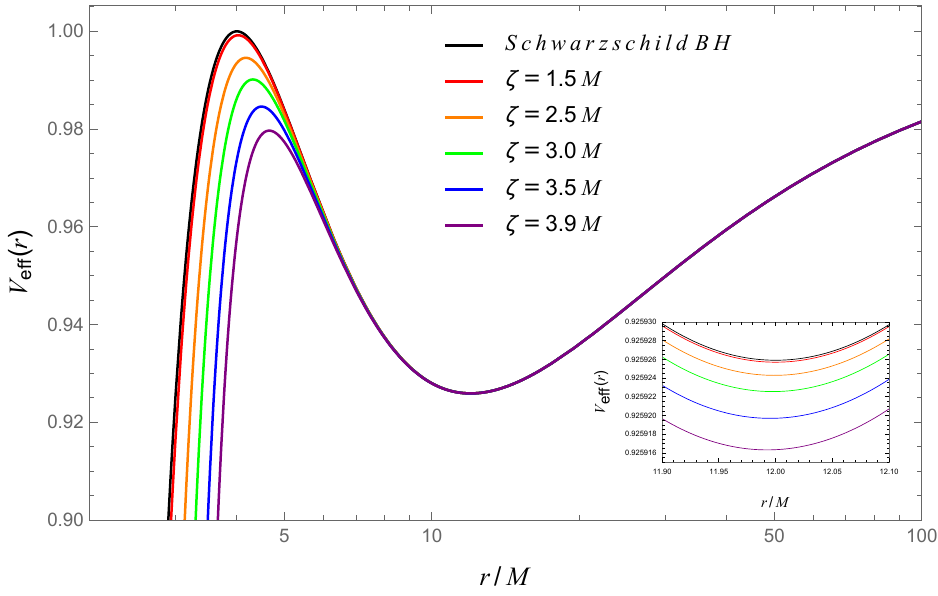}

		\caption{Comparison of the effective potential $V_{\text{eff}}(r)$ at $\tilde{L} = 4M$ between the QCBH (for $\zeta= 1.5M,2.5M, 3.0M,3.5M, 3.9M$) and a Schwarzschild BH of equal mass $M$.}
	\label{fig:Veff}
\end{figure}

Substituting the metric function from Eq.~\eqref{metric_function} into Eq.~\eqref{effective_potential}, we obtain the effective potential for the QCBH. This effective potential applies to particles orbiting the BH with a specific angular momentum $\tilde{L}$. Moreover, we can observe that since $\left. g_{tt} \right|_{r \rightarrow \infty} = 1$, it follows that $\left. V_{\rm{eff}} \right|_{r \rightarrow \infty} = 1$. Figure~\ref{fig:Veff} displays the effective potentials corresponding to different values of the parameter $\zeta$ for $\tilde{L}=4M$. It can be observed from the figure that as $\zeta$ approaches zero, the effective potential gradually approaches the Schwarzschild form.

\section{From bound orbits to periodic orbits}\label{sec:periodic}

\subsection{General features of bound orbits}

\begin{figure}[htbp]
	\centering
	\includegraphics[width=8cm]{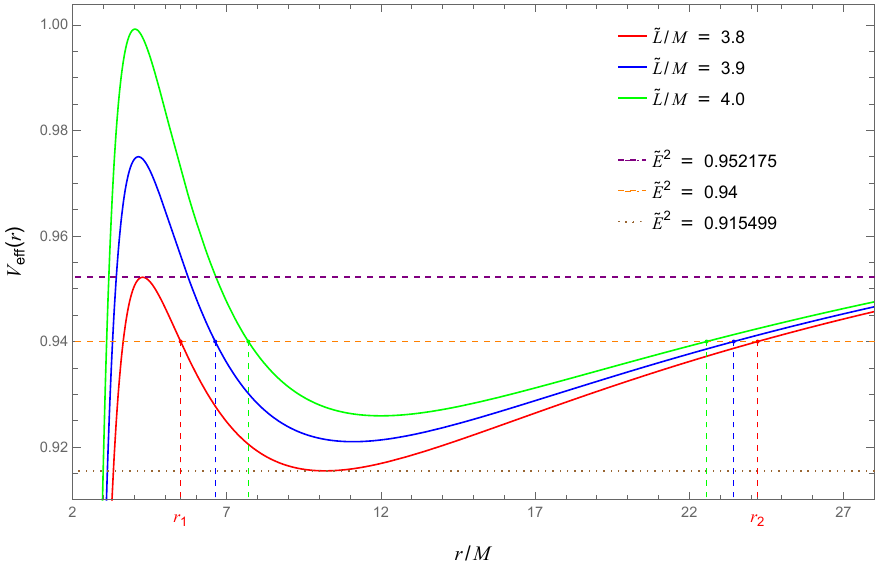}

	\caption{Relationship between the effective potential $V_{\rm{eff}}$ and the specific energy squared $\tilde{E}^2$ for different specific angular momenta. The vertical dashed lines mark the radial coordinate $r$ of the intersection points between $\tilde{E}^2$ and the effective potential curves, i.e., the locations of the turning points. The quantum parameter is chosen as $\zeta=1.5M$.}
	\label{fig:1_05_Veff_point}
\end{figure}

\begin{figure}[htbp]
	\centering
	\includegraphics[width=8cm]{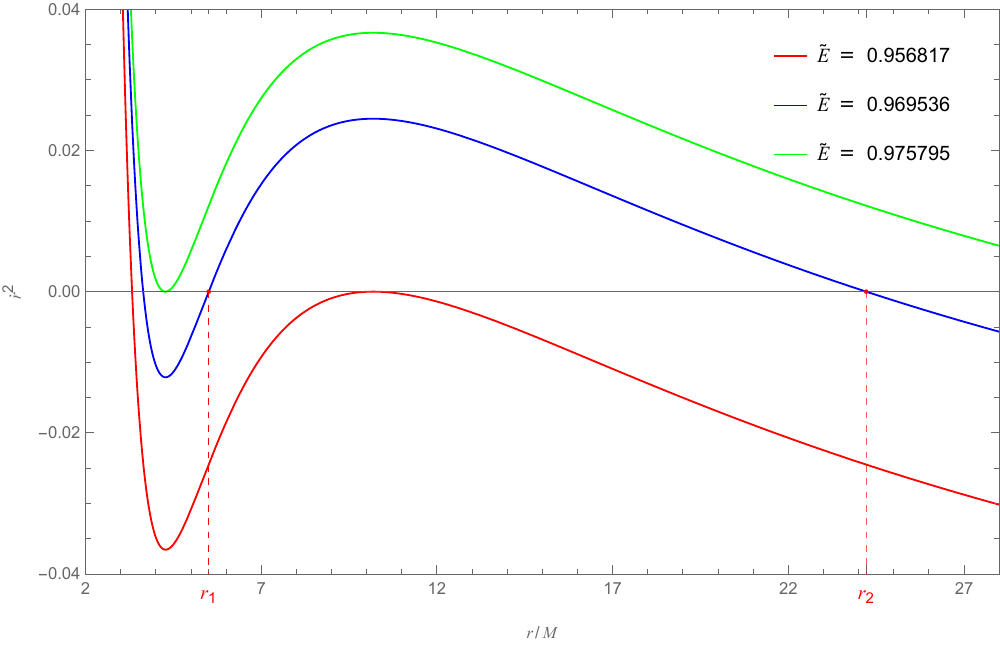}

	\caption{Dependence of $\dot{r}^2$ on $r/M$ for $\zeta = 1.5M$ and $\tilde{L}=3.8M$. The specific energy $\tilde{E}$ is set to $0.956817$, $0.969536$, and $0.975795$ from bottom to top. The vertical dashed lines mark the turning points.}
	\label{fig:r_dot_2}
\end{figure}

In general, bound orbits are characterized by two turning points, whose locations depend on the relationship between the particle's specific energy $\tilde{E}$ and the effective potential $V_{\rm{eff}}$. As an example, Fig.~\ref{fig:1_05_Veff_point} displays the relationship between $V_{\rm{eff}}$ and different values of $\tilde{E}$ for $\tilde{L}/M = 3.8$, $3.9$, and $4.0$. The three horizontal lines from bottom to top represent particle energies $\tilde{E}^2 = 0.915499$, $0.94$, and $0.952175$, corresponding to specific energies $\tilde{E}_1=0.956817$, $\tilde{E}_2=0.969536$, and $\tilde{E}_3=0.975795$, respectively. On one hand, for a fixed specific energy, the separation between the two turning points decreases as the specific angular momentum increases. On the other hand, if $\tilde{E}^2$ is less than the minimum of $V_{\rm{eff}}$, the particle will fall into the BH after a finite time of motion. Conversely, if $\tilde{E}^2$ is greater than the maximum of $V_{\rm{eff}}$, the final fate of the particle depends on its direction of motion; it will either fall into the BH or escape to infinity. Bound orbits are permitted only when $\tilde{E}^2$ lies between the minimum and maximum of the effective potential. Under these conditions, the two turning points of a bound orbit, denoted by $r_{1}$ and $r_{2}$, correspond to the intersections of the $\tilde{E}^2$ line with the effective potential curve. Therefore, for a fixed $\tilde{L}/M$, there exists an allowed range for $\tilde{E}$, and this range increases with $\tilde{L}/M$. For given $\tilde{E}$ and $\tilde{L}/M$, the turning points can be found by solving $V_{\rm{eff}}=\tilde{E}^2$. For instance, when $\tilde{E}^2 = 0.94$ and $\tilde{L}/M=3.8$, the turning points are located at $r_1 = 5.50549$ and $r_2=24.2157$. These conclusions can similarly be drawn from Fig.~\ref{fig:r_dot_2}.

\begin{figure*}[htbp]
	\centering
	\subfigure{\includegraphics[width=0.32\textwidth,height=3.8cm, keepaspectratio]{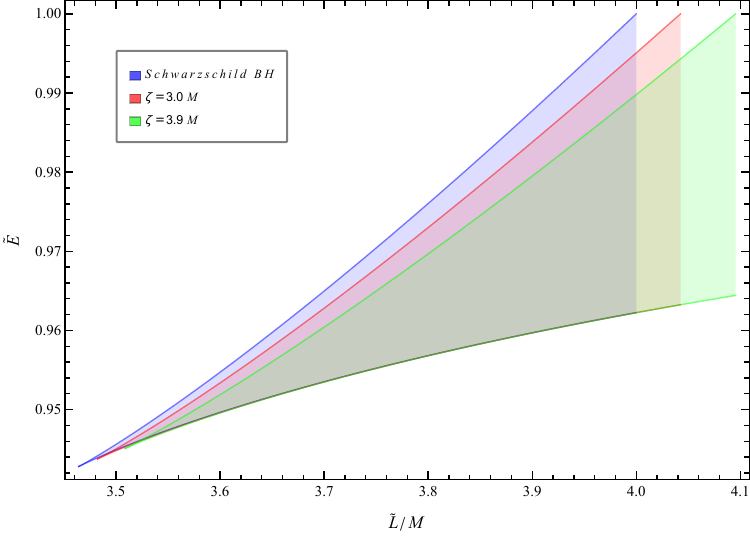}
		\includegraphics[width=0.32\textwidth,height=3.8cm, keepaspectratio]{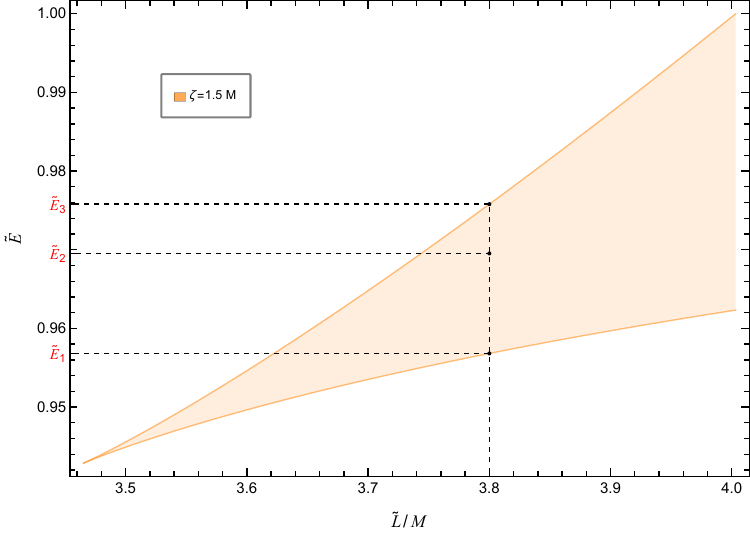}
 }

	\caption{Allowed parameter region in the $\tilde{E}$--$\tilde{L}$ plane for a massive particle moving around the QCBH. Left panel: $\zeta = 3.0M,\ 3.9M$, and the Schwarzschild case. Right panel: $\zeta = 1.5M$, with the vertical dashed line located at $\tilde{L}=3.8M$, while $\tilde{E}_1=0.956817$, $\tilde{E}_2=0.969536$, and $\tilde{E}_3=0.975795$, respectively.}
	\label{fig:E_L_Range}
\end{figure*}

Figure~\ref{fig:E_L_Range} presents the regions of bound orbits in the $\tilde{E}$-$\tilde{L}/M$ diagram for different $\zeta$. Bound orbits exist when the parameters fall within the shaded area; otherwise, no bound orbits are possible. Furthermore, for a fixed $\tilde{L}/M$, bound orbits correspond to a finite energy width, which increases with $\tilde{L}/M$, consistent with the above conclusion.

\begin{figure*}[htbp]
	\centering
	\subfigure{\includegraphics[width=0.32\textwidth,height=5.5cm, keepaspectratio]{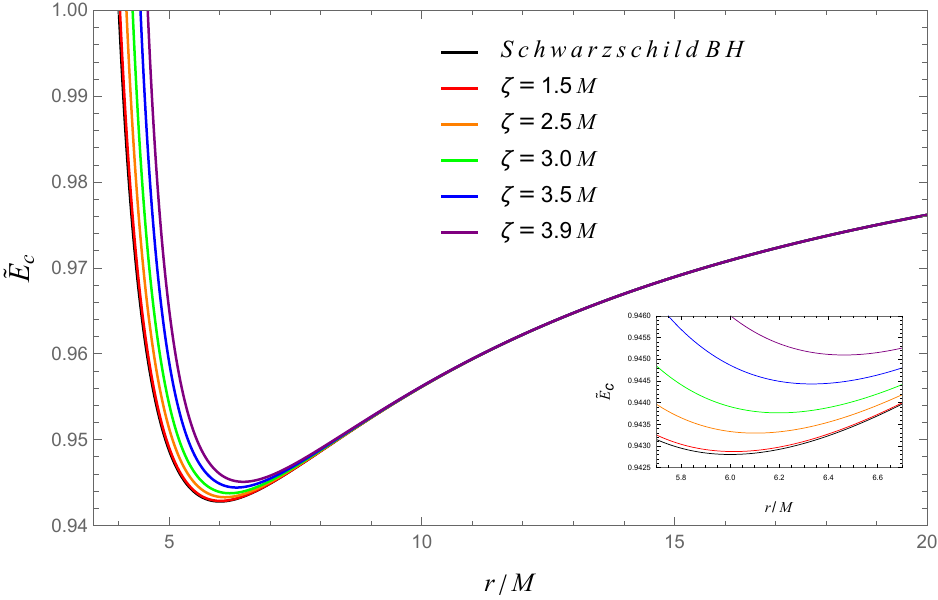}
		\includegraphics[width=0.32\textwidth,height=5.5cm, keepaspectratio]{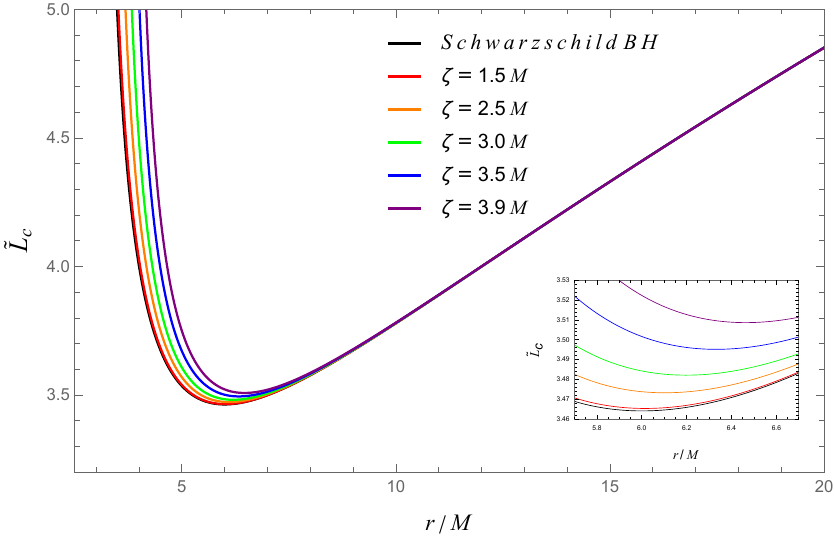}
		\includegraphics[width=0.32\textwidth,height=5.5cm, keepaspectratio]{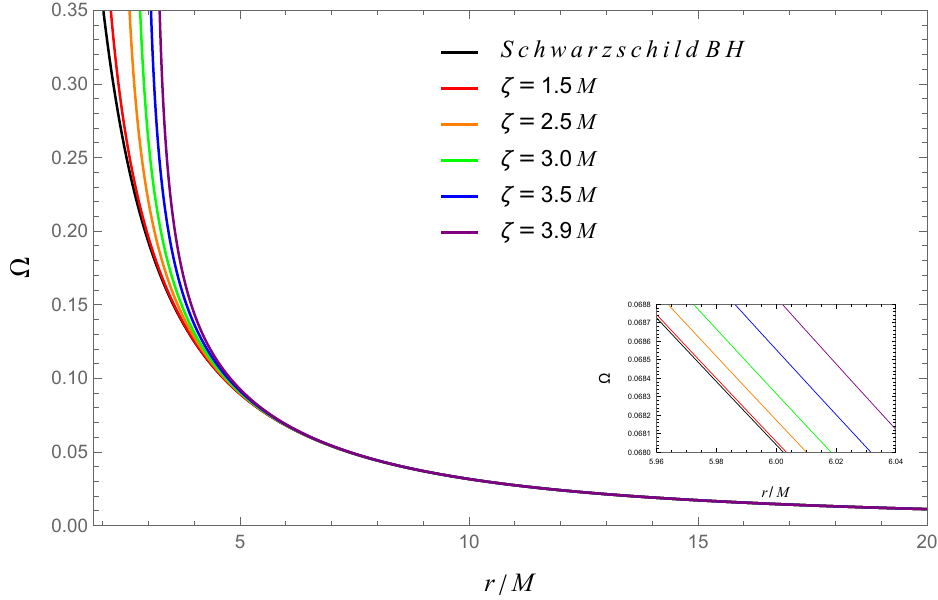}
 }

	\caption{Radial distribution curves of the specific energy (left panel), specific angular momentum (middle panel), and angular velocity (right panel) for a particle in circular motion under different values of $\zeta$.}
	\label{ELOmega}
\end{figure*}

\begin{figure*}[htbp]
	\centering
	\subfigure{\includegraphics[width=0.32\textwidth,height=3.8cm, keepaspectratio]{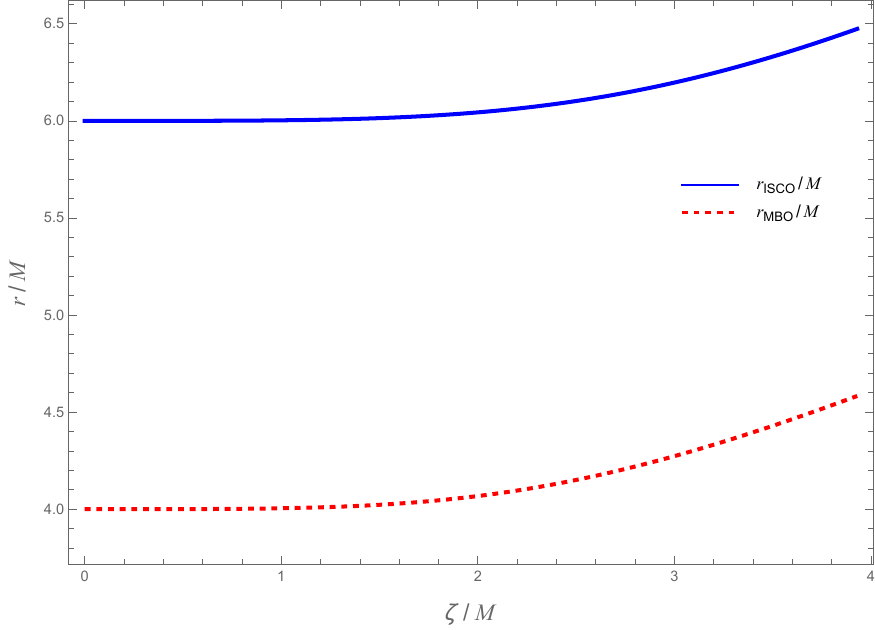}
		\includegraphics[width=0.32\textwidth,height=3.8cm, keepaspectratio]{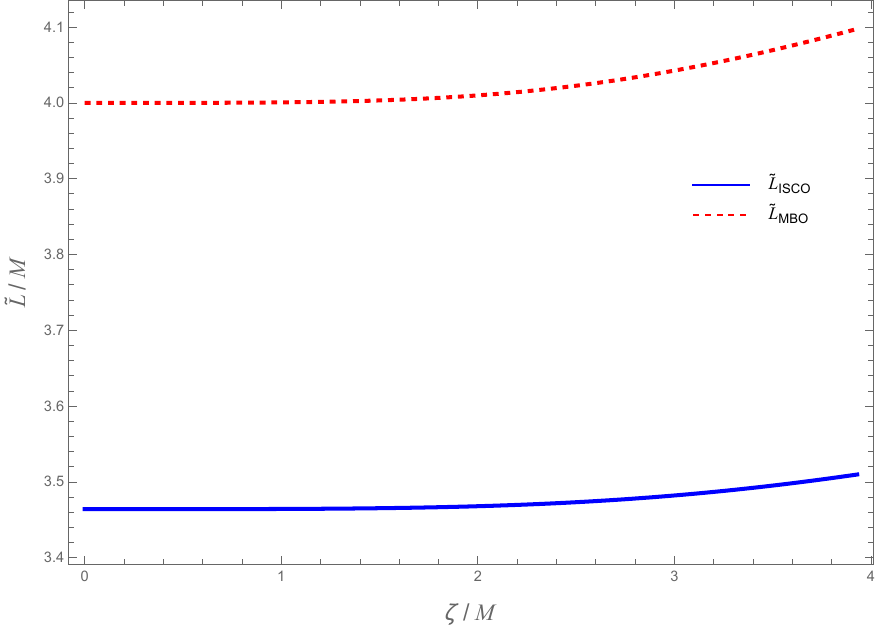}
		\includegraphics[width=0.32\textwidth,height=3.8cm, keepaspectratio]{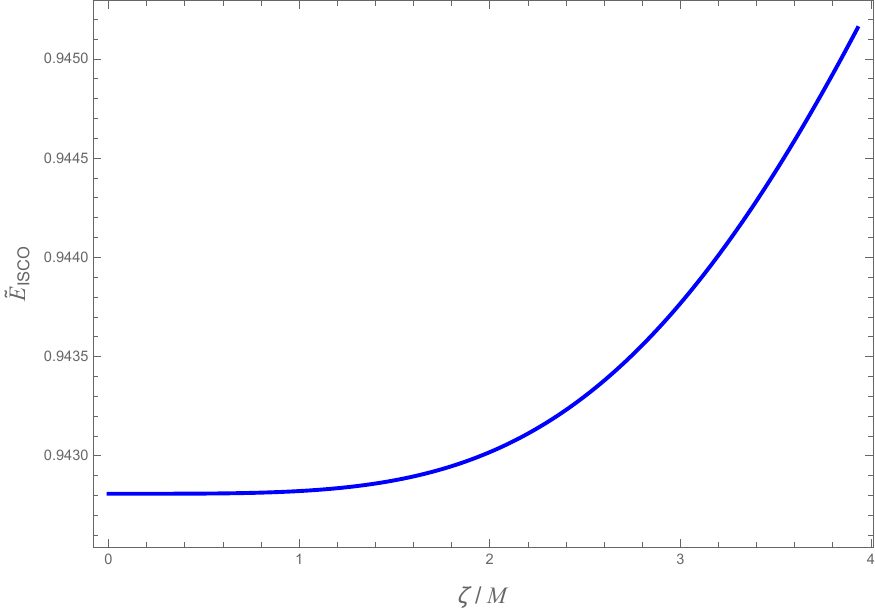}
 }

	\caption{The variation curves of $r_{\rm ISCO}$ and $r_{\rm MBO}$ with $\zeta$ (left panel), the variation curves of $L_{\rm ISCO}$ and $L_{\rm MBO}$ with $\zeta$ (middle panel), and the variation of $\tilde{E}_{\rm ISCO}$ with $\zeta$ (right panel).}
	\label{isco_mbo}
\end{figure*}

\begin{figure}[htbp]
	\centering
	\includegraphics[width=8cm]{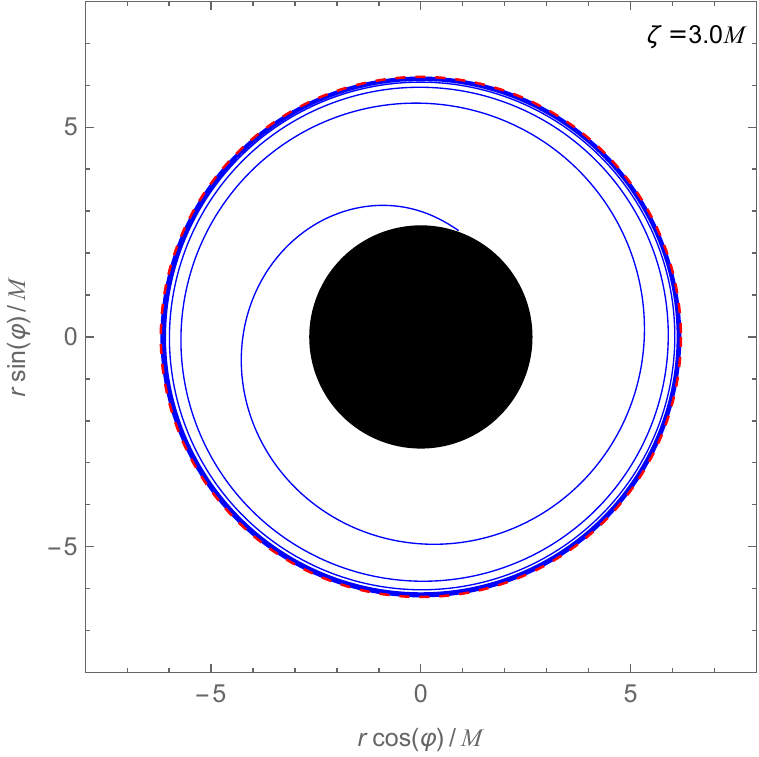}

	\caption{Orbital inspiral of a test particle from the BH ISCO to the event horizon of the QCBH with quantum parameter $\zeta=3.0M$.}
	\label{Inspirals}
\end{figure}

\subsection{Circular orbit}

Circular orbits in the equatorial plane require the simultaneous satisfaction of $\dot{r}=0$ and $\ddot{r}=0$, which implies that
\begin{equation}
 \begin{split}
 V_\text{eff}(r)= \tilde{E}^2, \quad
 \frac{{\rm d} V_{\rm eff}(r)}{{\rm d}r} = 0.\label{Circular_condition}
 \end{split}
\end{equation}
These constraints yield expressions for the specific energy $\tilde{E}_{\rm c}$ and specific angular momentum $\tilde{L}_{\rm c}$ on circular orbits
\begin{equation}
 \begin{split}
 \tilde{E}_{\rm c} = \frac{g_{tt}}{\sqrt{g_{tt} -\frac{1}{2}rg_{tt,r}}}, \\[12pt]
 \tilde{L}_{\rm c} = \sqrt{\frac{r^3g_{tt,r}}{2g_{tt} -rg_{tt,r}}}.
 \label{eq:EL}
 \end{split}
\end{equation}
Another important physical quantity on circular orbits, the angular velocity, is defined as~\cite{Page:1974he}
\begin{equation}
	\Omega=\frac{{\rm d}\varphi}{{\rm d}t}=\frac{\dot{\varphi}}{\dot{t}}=\sqrt{\frac{g_{tt,r}}{2r}}.\label{angular_velocity}
\end{equation}

The radial dependencies of $\tilde{E}$, $\tilde{L}$, and $\Omega$, along with their variation with the quantum parameter $\zeta$, are illustrated in Fig.~\ref{ELOmega}. It can be observed that, at a given radial distance $r$, a particle orbiting the QCBH possesses larger $\tilde{E}$, $\tilde{L}$, and $\Omega$ than one orbiting the Schwarzschild BH. Moreover, $\Omega$ decreases as $r$ increases. In contrast, $\tilde{E}$ and $\tilde{L}$ exhibit a different behavior: as $r$ increases, these two quantities first decrease and then increase.

\subsection{Marginally bound orbit and ISCO}

There are two special bound circular orbits, referred to as the marginally bound orbit (MBO) and the ISCO. In some literature, the MBO is also called the innermost bound circular orbit (IBCO)~\cite{Misra:2010pu,Liu:2018vea}. First, we consider the MBO, which corresponds to the critical bound state where the particle's specific energy reaches $\tilde{E} = 1$, causing the outer turning point to tend to infinity. Beyond this value, the particle would possess a positive radial velocity at infinity, i.e., $\dot{r}^{2} =\left. \Bigl( 1 - \frac{4M^{2}\zeta^{4}}{r^{6}} \Bigr) \Bigl[ \tilde{E}^{2} - V_{\rm{eff}} \Bigr] \right|_{r \rightarrow \infty} > 0$, and no bound orbit can exist. Therefore, $\tilde{E} = 1$ represents the maximum energy admissible for bound orbital motion. Consequently, in addition to satisfying Eq.~\eqref{Circular_condition}, the MBO must further satisfy~\cite{Deng:2020yfm,Tu:2023xab}
\begin{equation}
	V_{\rm eff}(r)= 1. \label{MBO_condition}
\end{equation}
Combining the first equation in Eq.~\eqref{eq:EL} and Eq.~\eqref{MBO_condition}, we get
\begin{equation}
	r_{\rm MBO} = \frac{2\left[g_{tt}(r_{\rm MBO})-g_{tt}(r_{\rm MBO})^2\right]}{g_{tt,r}(r_{\rm MBO})}.\label{eq:rMBO}
\end{equation}
Then, combining the second equation in Eq.~\eqref{eq:EL}, we obtain the expression for $\tilde{L}_{MBO}$ as
\begin{equation}
 \tilde{L}_{\rm MBO} = \frac{2\sqrt{g_{tt}(r_{\rm MBO})\left[1-g_{tt}(r_{\rm MBO})^3\right]}}{g_{tt,r}(r_{\rm MBO})}.\label{eq:LMBO}
\end{equation}
The radius $r_{\rm MBO}$ and special angular momentum $\tilde{L}_{MBO}$ for a fixed $\zeta$ are obtained by solving Eqs.~\eqref{eq:rMBO} and \eqref{eq:LMBO}. For the extremal BH case where $\zeta = 3.9M$, the values are $r_{\rm MBO} = 4.5718M$ and $\tilde{L}_{MBO}= 4.0955M$.

Then, for the ISCO of the BH, this defines the minimum radius for which a stable circular orbit is possible. Any circular orbit lying inside the ISCO is unstable, and a perturbed particle on such an orbit will inevitably spiral inward until it is captured~\cite{Mummery:2022ana,Ko:2023igf}, as shown in Fig.~\ref{Inspirals}. Therefore, in addition to satisfying Eq.~\eqref{Circular_condition}, the ISCO also satisfies~\cite{Deng:2020yfm,Tu:2023xab}
\begin{equation}
	\left. \frac{{\rm d}^{2}V_{\rm eff}(r)}{{{\rm d}r}^{2}} \right|_{r = r_{\rm ISCO}} = 0,\label{eq:second_condition}
\end{equation}
which leads to the result
\begin{equation}
	r_{\rm ISCO}=\frac{3g_{tt}(r_{\rm ISCO})g_{tt,r}(r_{\rm ISCO})}{2g_{tt,r}(r_{\rm ISCO})^2-g_{tt}(r_{\rm ISCO})g_{tt,rr}(r_{\rm ISCO})}.
	\label{eq:isco}
\end{equation}
Substituting the result from the above calculation into Eq.~\eqref{eq:EL} yields $\tilde{E}_{\rm ISCO}$ and $\tilde{L}_{\rm ISCO}$. Figure~\ref{isco_mbo} shows the numerical results for $\tilde{r}_{\rm MBO}$, $\tilde{r}_{\rm ISCO}$, $\tilde{L}_{\rm MBO}$, $\tilde{L}_{\rm ISCO}$, and $\tilde{E}_{\rm ISCO}$. The figure clearly reveals an increasing trend in these quantities as $\zeta$ increases. Meanwhile, it can also be confirmed by direct calculation that among all circular orbits, the specific energy and specific angular momentum of a particle reach their minimum values at $r_{\rm{isco}}$~\cite{Konoplya:2009ig}.

\subsection{Periodic orbits}

Circular orbits, discussed in the previous section, represent the simplest bound orbits. More generally, bound orbits can be classified as periodic when they exactly close after a finite number of oscillations. Indeed, any generic bound orbit can be viewed as a perturbation of a periodic one~\cite{Levin:2008mq}. Therefore, analyzing periodic orbits provides significant insight into the behavior of generic orbits. In this section, we focus on periodic orbits and their characterization, which will serve as the foundation for generating gravitational waveforms in the next section.

In the spherically symmetric BH background considered here, such an orbit is fully characterized by the oscillation frequencies in the radial and azimuthal directions i.e., $\omega_{r}$ and $\omega_{\varphi}$. Specifically, for an orbit to be periodic, the ratio of these two frequencies must be a rational number, ensuring that the particle returns to its starting point after a finite time interval.

Considering a bound orbit with turning points $r_1$ and $r_2$, the particle oscillates between these radii. The apsidal angle $\Delta\varphi$ swept per radial period is given by
\begin{equation}
	\Delta\varphi =2 \oint \rm d\varphi, \label{Delta_varphi}
\end{equation}
The factor of 2 arises from the symmetry of the particle trajectory.

Following Ref.~\cite{Levin:2008mq}, we define a parameter
\begin{equation}
	q\equiv \frac{\omega_\varphi}{\omega_r}-1 = \frac{\Delta\varphi}{2\pi} - 1, \label{q}
\end{equation}
Using the trajectory equation and combining with Eq.~\eqref{Delta_varphi}, it can be expressed as
\begin{equation}
	q=\frac{1}{\pi} \int_{r_1}^{r_2} \frac{1}{r^2 \sqrt{\left( \frac{\tilde{E} ^2}{\tilde{L}^2}-\frac{g_{\rm tt}}{\tilde{L}^2}-\frac{g_{\rm tt}}{r^2}\right) \frac{g_{\rm rr}}{g_{\rm tt}}}}\,{\rm d}r-1. \label{q_varphi}
\end{equation}
Equation~\eqref{q_varphi} shows that $q$ depends on the specific energy $\tilde{E}$, specific angular momentum $\tilde{L}$, and the metric functions $g_{\rm tt}$, $g_{\rm rr}$. Consequently, $q$ varies with the parameter $\zeta$ characterizing the BH.

\begin{figure}[htbp]
	\centering
	\includegraphics[width=8cm]{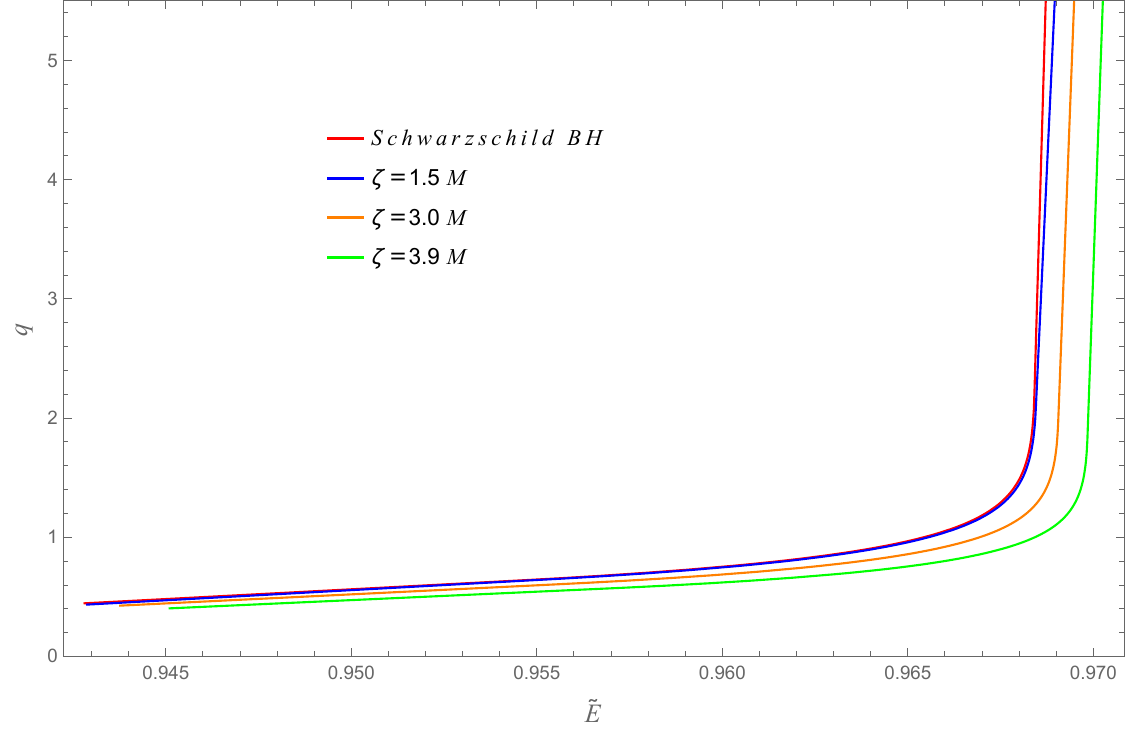}

	\caption{Rational number $q$ versus $\tilde{E}$ for different values of $\zeta$, with the specific angular momentum fixed at $\tilde{L}=(\tilde{L}_{\rm ISCO}+\tilde{L}_{\rm MBO})/2$.}
	\label{q_E}
\end{figure}

\begin{figure}[htbp]
	\centering
	\includegraphics[width=8cm]{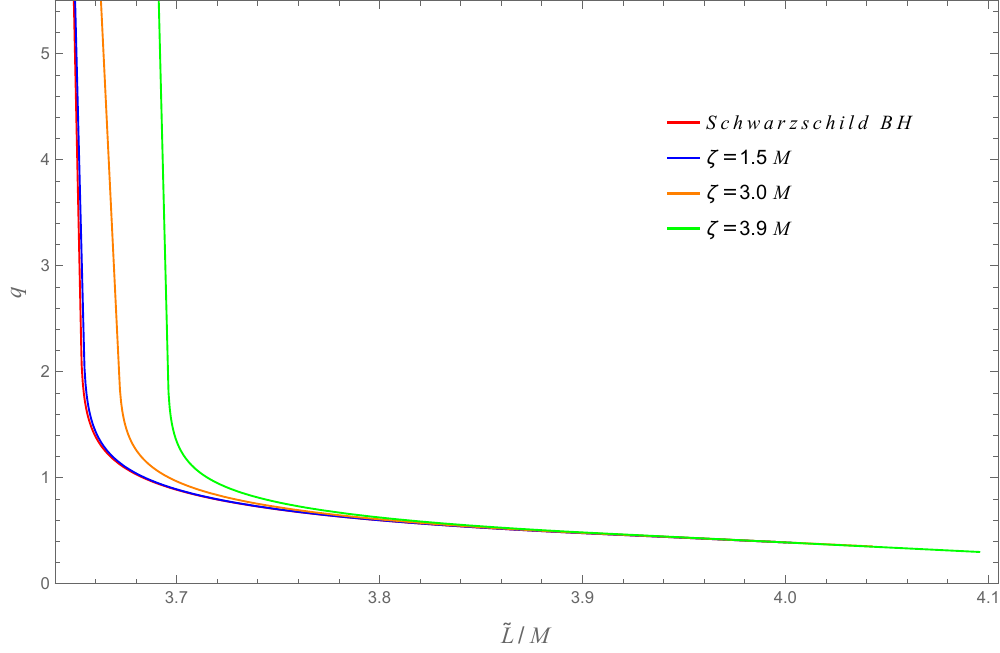}

	\caption{Rational number $q$ versus $\tilde{L}$ for different values of $\zeta$, with the specific energy fixed at $\tilde{E}=0.96$.}
	\label{q_L}
\end{figure}

For a bound orbit with given $\zeta$, $\tilde{E}$, and $\tilde{L}$, one can compute $q$. Its behavior is illustrated in Figs.~\ref{q_E} and \ref{q_L}, which plot $q$ against $\tilde{E}$ and $\tilde{L}$, respectively. The specific angular momentum for a bound orbit lies between $\tilde{L}_{\rm ISCO}$ and $\tilde{L}_{\rm MBO}$. Figure~\ref{q_E} displays $q$ as a function of $\tilde{E}$ for the fixed specific angular momentum $\tilde{L}_{\rm fixed}=(\tilde{L}_{\rm ISCO}+\tilde{L}_{\rm MBO})/2$. Initially, $q$ increases slowly with $\tilde{E}$, but diverges sharply as $\tilde{E}$ approaches a maximum. This maximum energy value increases with $\zeta$.
Figure~\ref{q_L} shows $q$ versus $\tilde{L}/M$ for the fixed specific energy $\tilde{E}=0.96$. In all cases, $q$ decreases with increasing $\tilde{L}$ and diverges at the minimum allowed $\tilde{L}$. This minimum $\tilde{L}$ increases with $\zeta$.
As mentioned earlier, for a periodic orbit, $q$ must be a rational number. Therefore it is characterized by the three integers $(z, w, v)$ as
\begin{equation}
	q = w + \frac{v}{z}. \label{32}
\end{equation}
Prior studies~\cite{Levin:2008mq,Levin:2008ci} have used these integers to classify periodic orbits around various compact objects. Each integer has a geometric interpretation: $z$ (zoom number), $w$ (whirl number), and $v$ (vertex number) correspond, respectively, to the number of leaves, the number of whirls, and the number of vertices formed by joining successive apastra~\cite{Misra:2010pu,Babar:2017gsg,Chen:2025aqh}.

Table~\ref{tab:E_zwv} lists the fixed specific angular momentum $\tilde{L}_{\rm fixed}=(\tilde{L}_{\rm ISCO}+\tilde{L}_{\rm MBO})/2$ for different $\zeta$ values and the corresponding $\tilde{E}_{(z,w,v)}$ values for periodic orbits with specific $(z,w,v)$ parameters. The table shows that the specific energy is lowest for the Schwarzschild case ($\zeta =0$) and increases with $\zeta$. Table~\ref{tab:L_zwv} lists the $\tilde{L}_{(z,w,v)}$ values for periodic orbits with given $(z,w,v)$ parameters under the fixed specific energy $\tilde{E}=0.96$ for different $\zeta$ values. Similarly, the specific angular momentum is lowest for the Schwarzschild case ($\zeta =0$) and increases with $\zeta$.

\begin{figure*}[htbp]
 \centering
 \includegraphics[width=0.32\textwidth,height=5cm,keepaspectratio]{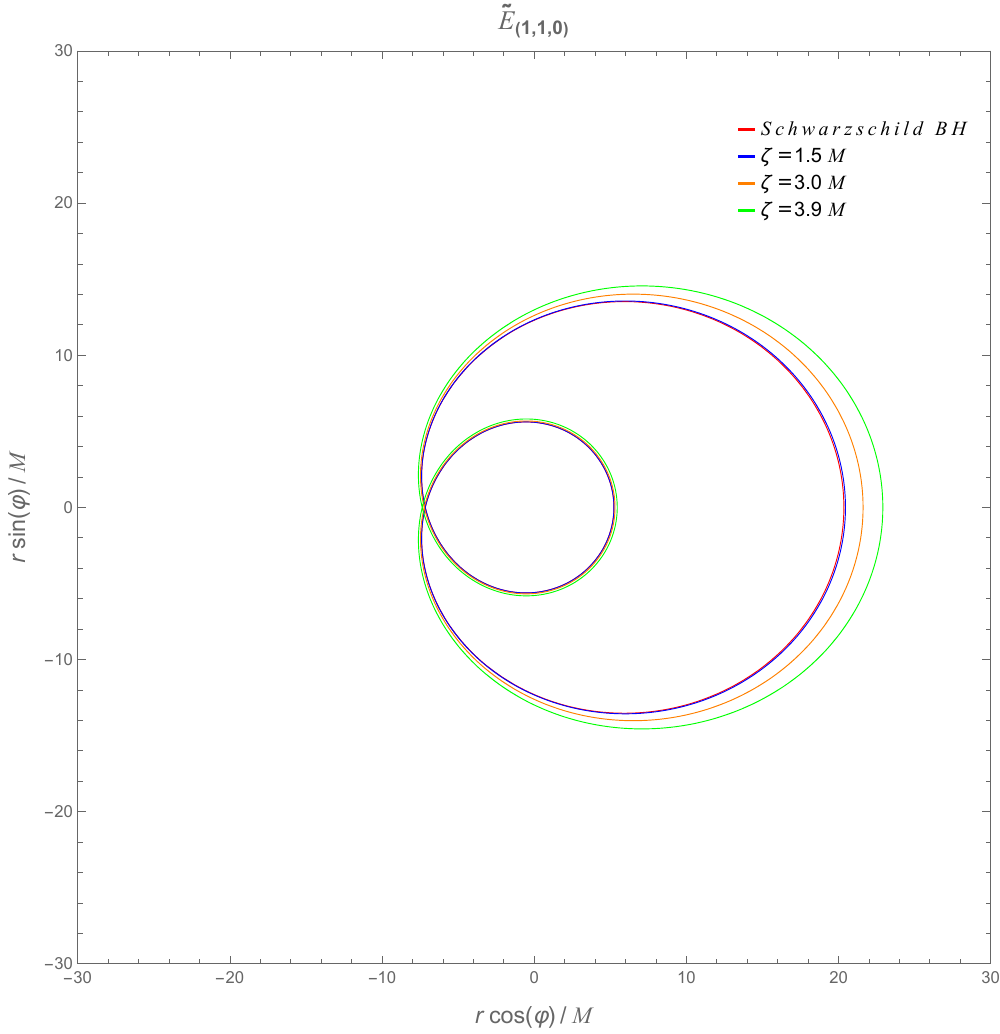}
 \hspace{0.5cm}
 \includegraphics[width=0.32\textwidth,height=5cm,keepaspectratio]{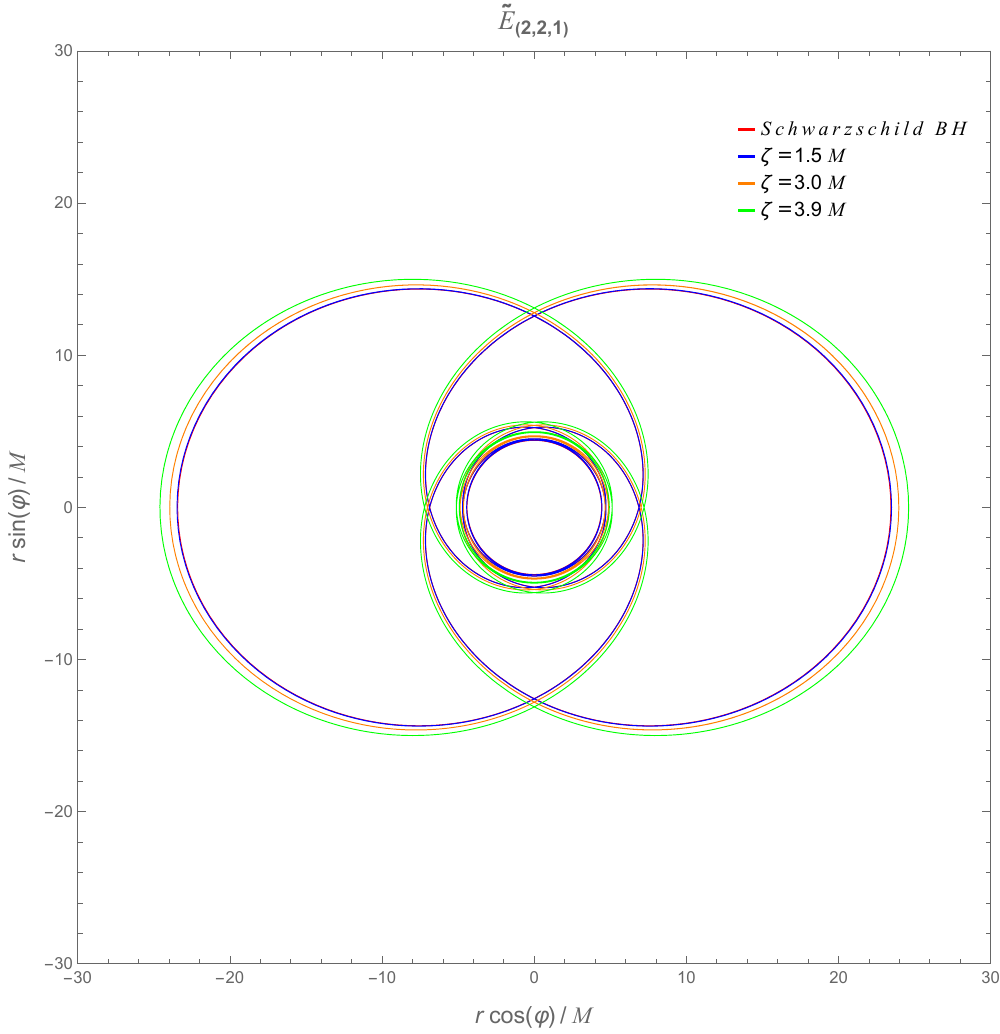}
 \hspace{0.5cm}
 \includegraphics[width=0.32\textwidth,height=5cm,keepaspectratio]{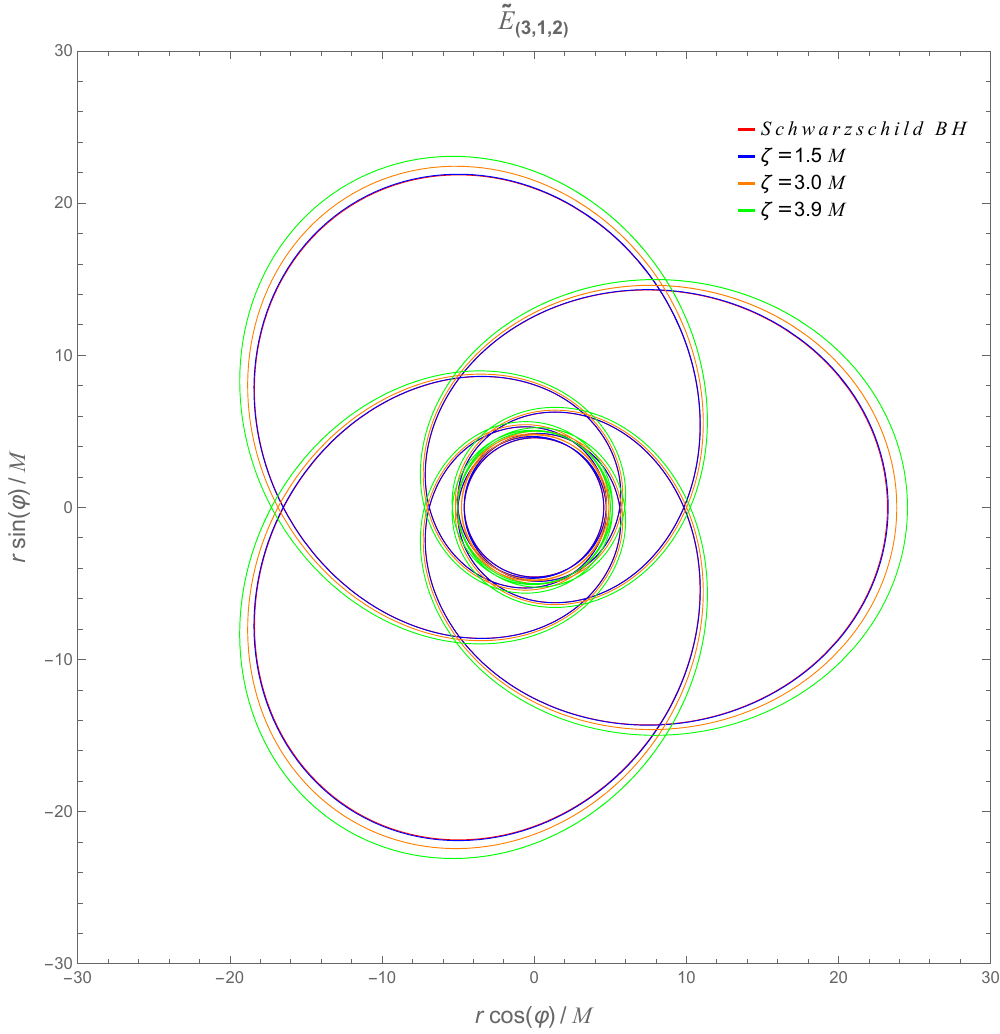}

 \vspace{0.5em}

 \includegraphics[width=0.32\textwidth,height=5cm,keepaspectratio]{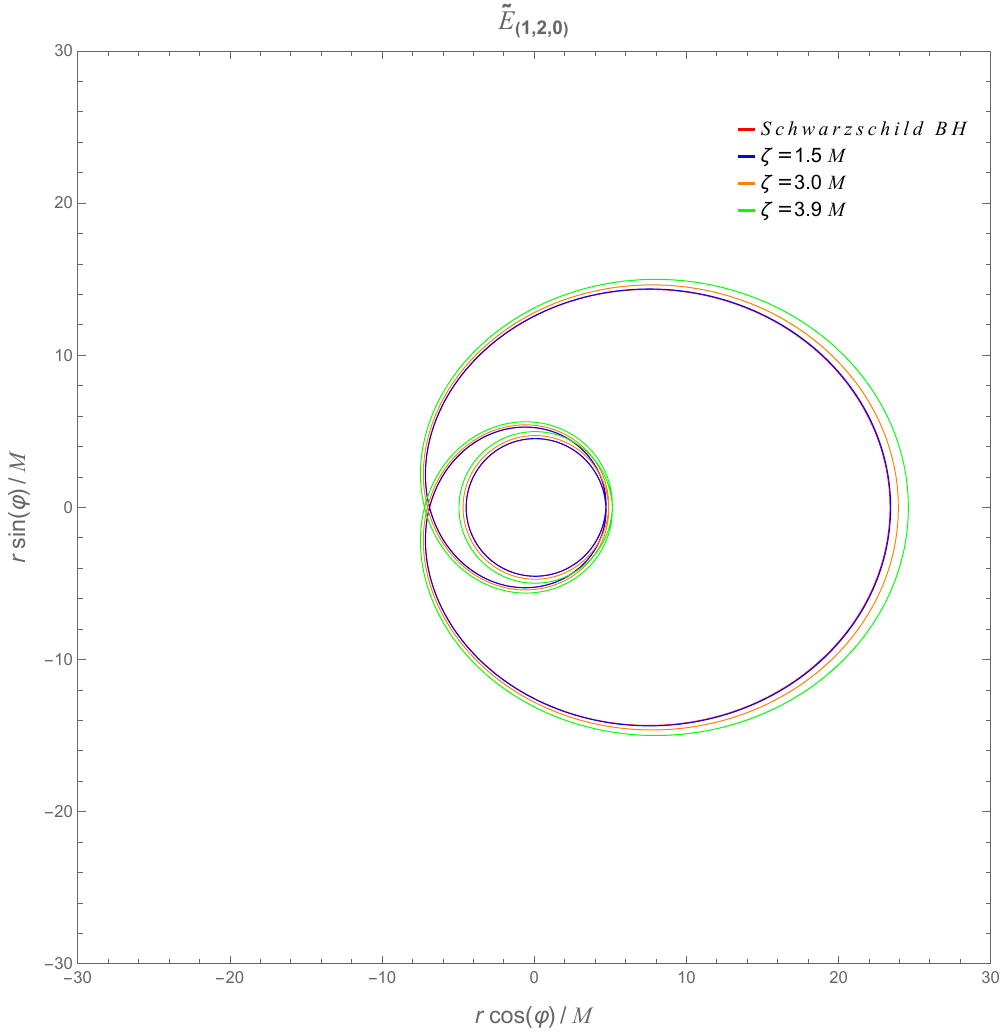}
 \hspace{0.5cm}
 \includegraphics[width=0.32\textwidth,height=5cm,keepaspectratio]{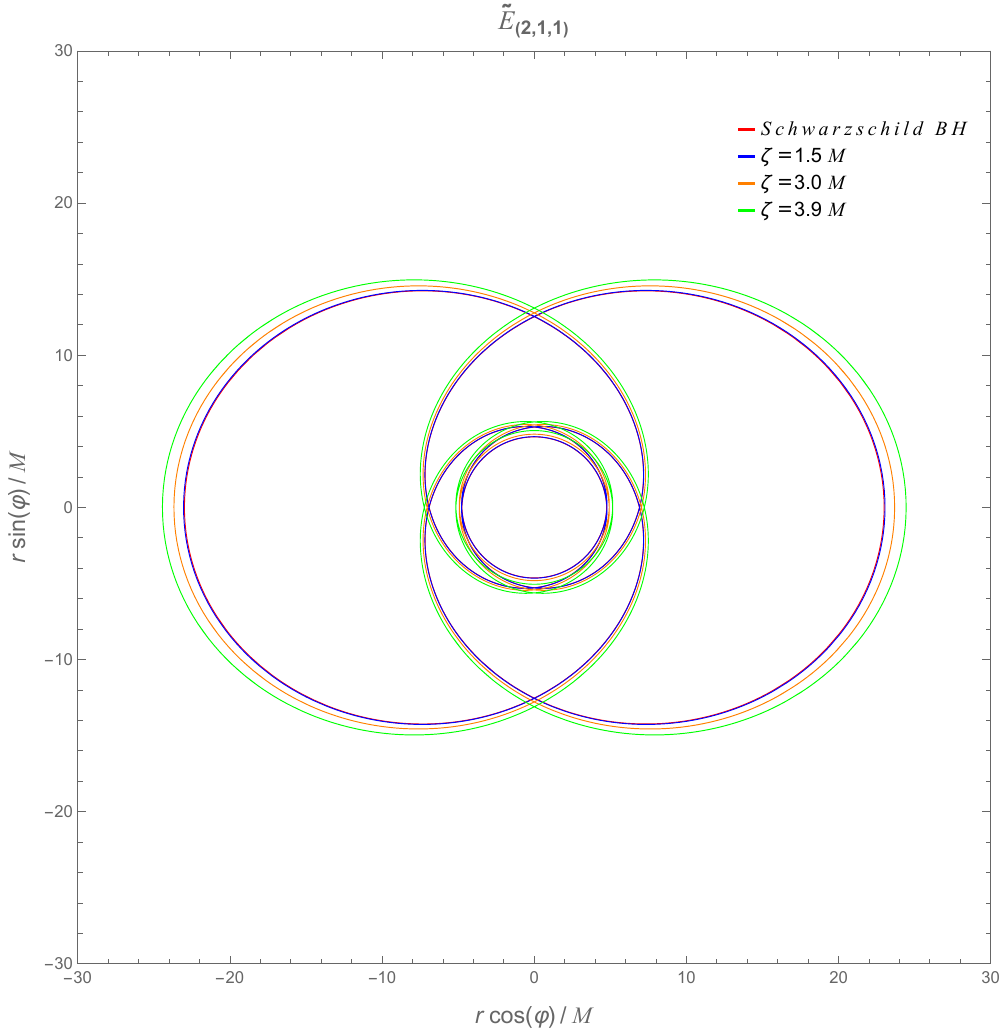}
 \hspace{0.5cm}
 \includegraphics[width=0.32\textwidth,height=5cm,keepaspectratio]{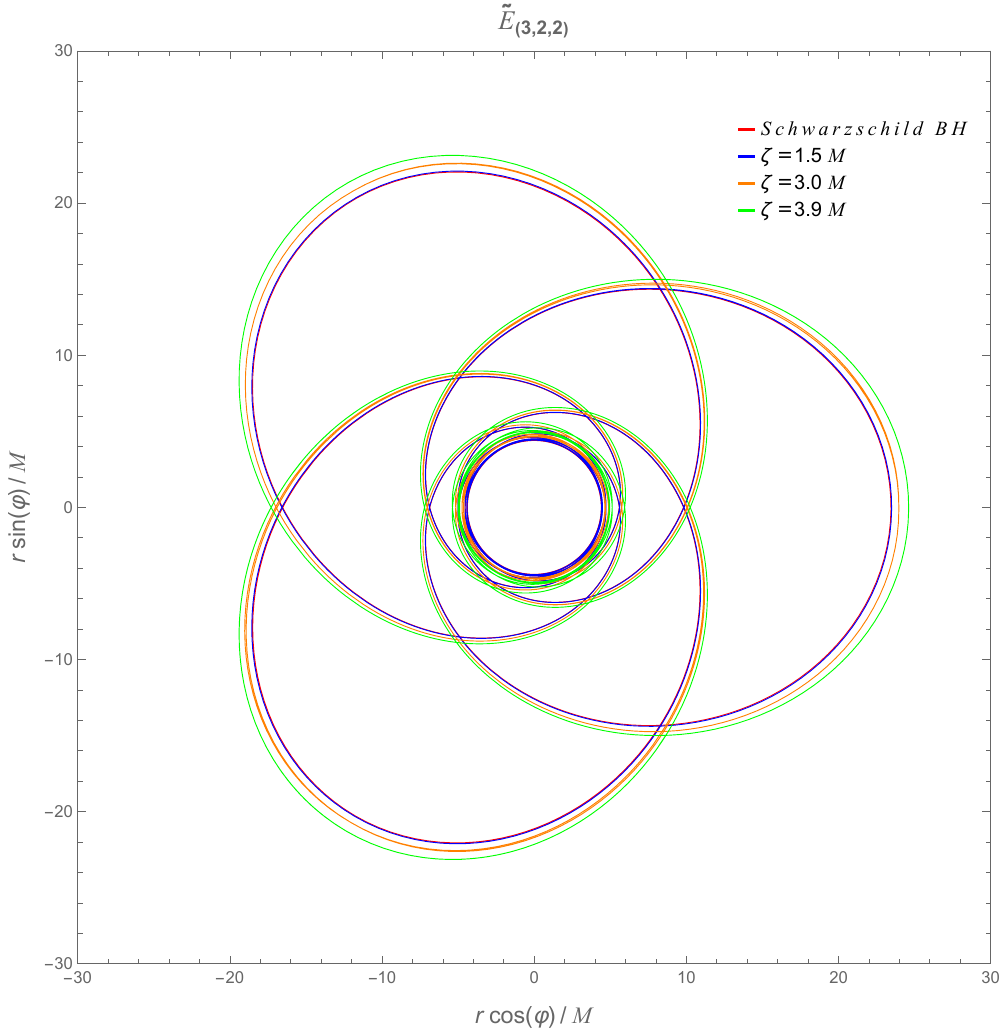}

	\caption{Periodic orbits corresponding to different integer sets $(z, w, v)$ around the QCBH. The parameter values are set to $\zeta/M = 0, 1.5, 3.0, 3.9$ and $\tilde{L}_{\rm fixed} =(\tilde{L}_{\rm ISCO} + \tilde{L}_{\rm MBO})/2$}
	\label{Orbits_E}
\end{figure*}

\begin{figure*}[htbp]
	\centering
	\includegraphics[width=0.32\textwidth,height=5cm,keepaspectratio]{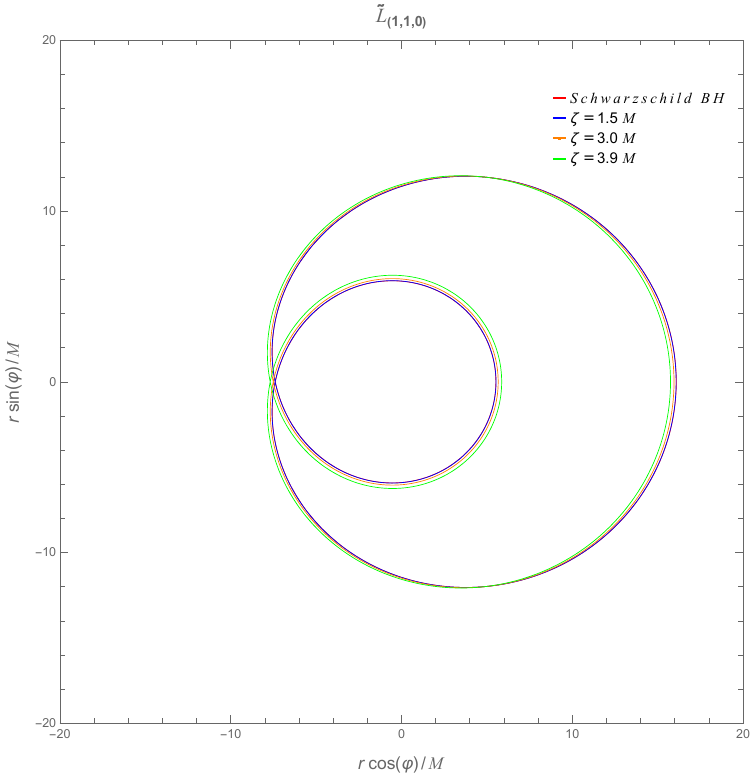}
	\hspace{0.5cm}
	\includegraphics[width=0.32\textwidth,height=5cm,keepaspectratio]{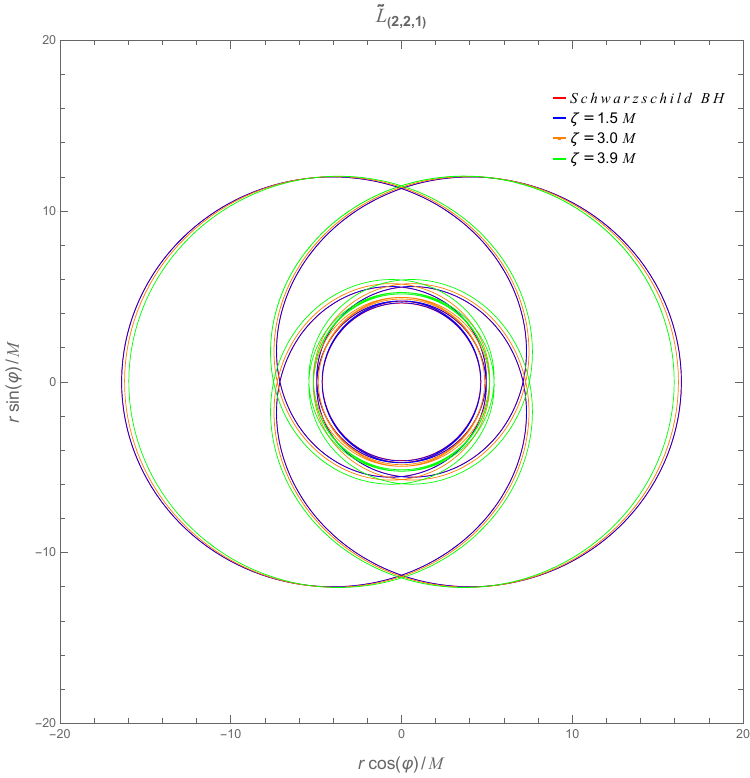}
	\hspace{0.5cm}
	\includegraphics[width=0.32\textwidth,height=5cm,keepaspectratio]{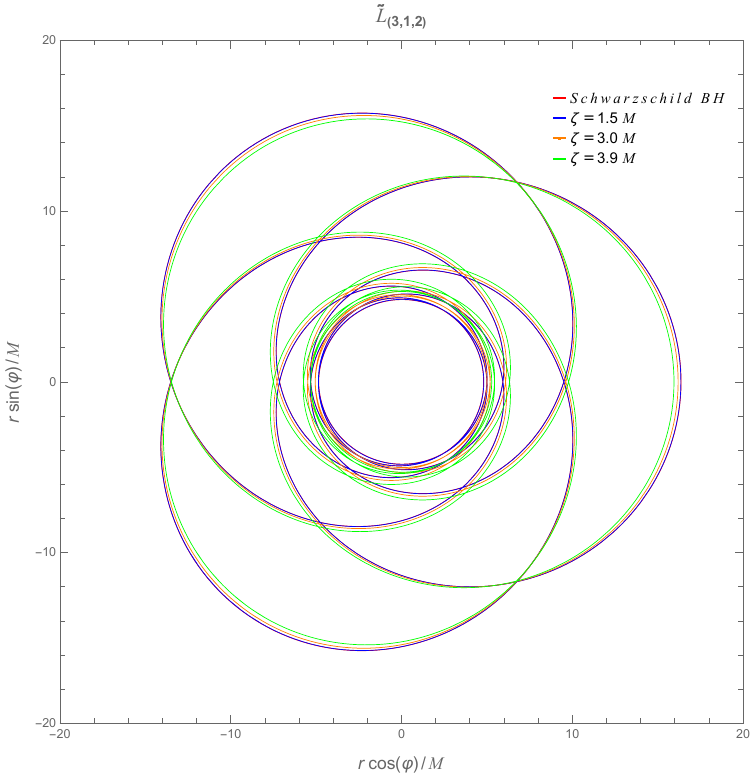}

	\vspace{0.5em}

	\includegraphics[width=0.32\textwidth,height=5cm,keepaspectratio]{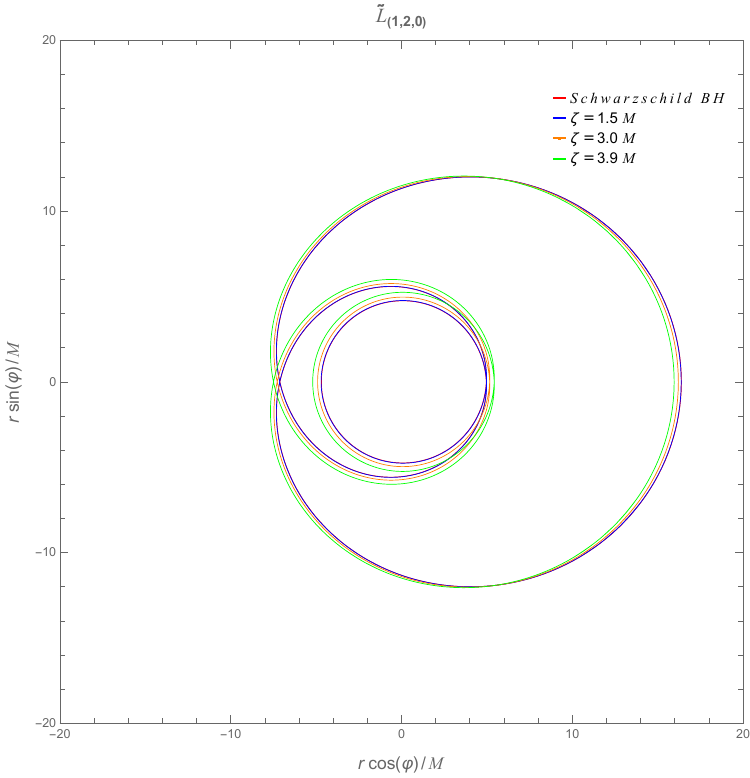}
	\hspace{0.5cm}
	\includegraphics[width=0.32\textwidth,height=5cm,keepaspectratio]{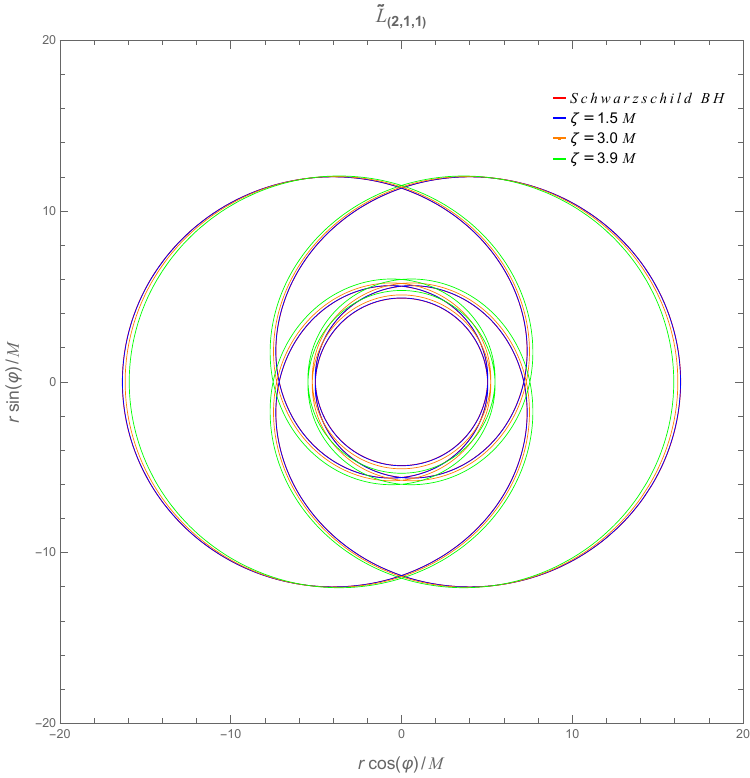}
	\hspace{0.5cm}
	\includegraphics[width=0.32\textwidth,height=5cm,keepaspectratio]{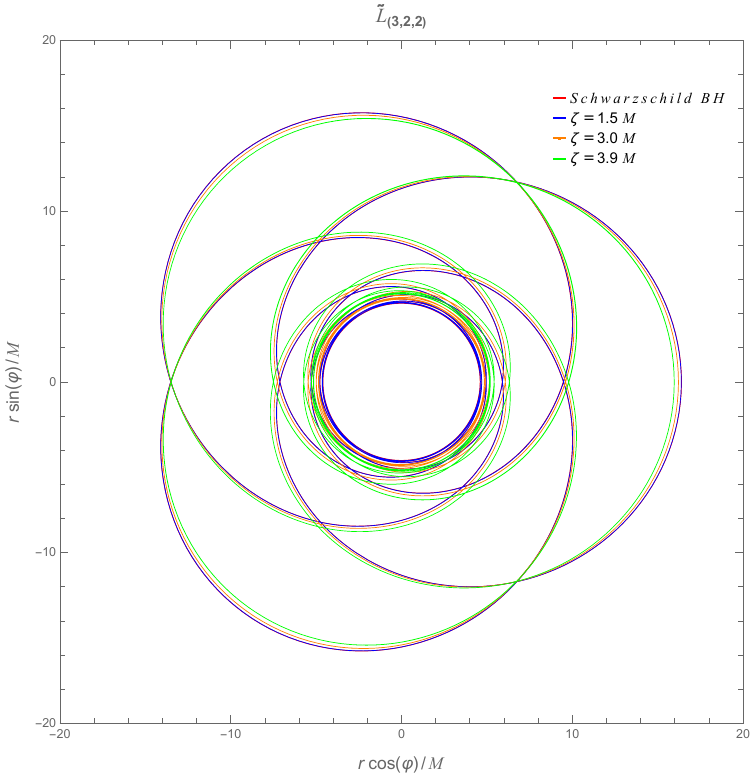}

	\caption{Periodic orbits corresponding to different integer sets $(z, w, v)$ around the QCBH. The parameter values are set to $\zeta/M = 0, 1.5, 3.0, 3.9$ and $\tilde{E}=0.96$.}
	\label{Orbits_L}
\end{figure*}

The periodic orbits are constructed by numerically integrating Eq.~\eqref{eq:motion_equation} for the $\tilde{E}_{(z,w,v)}$ and $\tilde{L}_{(z,w,v)}$ values listed in Table~\ref{tab:E_zwv} and Table~\ref{tab:L_zwv}, yielding the dependence of $r$ on $\varphi$. For the fixed specific angular momentum $\tilde{L}_{\rm fixed}$, Fig.~\ref{Orbits_E} shows examples of orbits with different integer sets $\tilde{L}_{(z,w,v)}$ in the QCBH. For the fixed specific energy $\tilde{E} = 0.96$, Fig.~\ref{Orbits_L} displays examples of orbits with different $\tilde{L}_{(z,w,v)}$ values in the QCBH. Inspection of these figures reveals that, as the quantum parameter $\zeta$ increases, the particle orbits in the QCBH spacetime background deviate more significantly from the Schwarzschild BH case.

\begin{table*}[htbp]

	\caption{The specific energy $\tilde{E}_{(z,w,v)}$ for QCBH with different values of $\zeta$ and a fixed $\tilde{L}_{\rm fixed} =(L_{\rm ISCO} + L_{\rm MBO})/2$.}
	\setlength{\heavyrulewidth}{0.08em}
	\setlength{\lightrulewidth}{0.14em}
	\setlength{\tabcolsep}{12.5pt}
	\begin{tabular}{ccccccccccc}
		\toprule
		&{\textbf{BHs}}
		&$\tilde{E}_{\scriptscriptstyle(1,1,0)}$ &$\tilde{E}_{\scriptscriptstyle(1,2,0)}$
		&$\tilde{E}_{\scriptscriptstyle(2,1,1)}$ &$\tilde{E}_{\scriptscriptstyle(2,2,1)}$
		&$\tilde{E}_{\scriptscriptstyle(3,1,2)}$ &$\tilde{E}_{\scriptscriptstyle(3,2,2)}$ &$\tilde{L}_{\rm fixed}$\\
		\midrule
		&{\textbf{Schwarzschild BH}} &0.965425 & 0.968383 & 0.968026 & 0.968434 & 0.968225 & 0.968438 & 3.732051\\
		& QCBH ($\zeta=1.5M$) &0.965552 & 0.968437 & 0.968095 & 0.968486 & 0.968286 & 0.968490 & 3.734293\\
		& QCBH ($\zeta=3.0M$) &0.966932 & 0.969060 & 0.968845 & 0.969086 & 0.968969 & 0.969087 & 3.762360\\
		& QCBH ($\zeta=3.9M$) &0.968421 & 0.969860 & 0.969737 & 0.969873 & 0.969810 & 0.969874 & 3.802094\\
		\bottomrule
		\label{tab:E_zwv}
	\end{tabular}
\end{table*}

\begin{table*}[htbp]

 \caption{The specific angular momentum $\tilde{L}_{(z,w,v)}$ for QCBH with different values of $\zeta$, with $\tilde{E}=0.96$ fixed.}
 \setlength{\heavyrulewidth}{0.08em} 
 \setlength{\lightrulewidth}{0.14em} 
 \setlength{\tabcolsep}{16pt}
 \begin{tabular}{cccccccccc}
 \toprule
 &{\textbf{BHs}}
 &$\tilde{L}_{\scriptscriptstyle(1,1,0)}$&$\tilde{L}_{\scriptscriptstyle(1,2,0)}$ &$\tilde{L}_{\scriptscriptstyle(2,1,1)}$
 &$\tilde{L}_{\scriptscriptstyle(2,2,1)}$ &$\tilde{L}_{\scriptscriptstyle(3,1,2)}$ &$\tilde{L}_{\scriptscriptstyle(3,2,2)}$ \\
 \midrule
 &{\textbf{Schwarzschild BH}} & 3.683588 & 3.653406 & 3.657596 & 3.652700 & 3.655335 & 3.652636 \\
 & QCBH ($\zeta=1.5M$) & 3.684466 & 3.654733 & 3.658818 & 3.654054 & 3.656608 & 3.653990 \\
 & QCBH ($\zeta=3.0M$) & 3.696412 & 3.671632 & 3.674671 & 3.671191 & 3.672982 & 3.671154 \\
 & QCBH ($\zeta=3.9M$) & 3.715303 & 3.695647 & 3.697797 & 3.695376 & 3.696572 & 3.695357\\
 \bottomrule
 \label{tab:L_zwv}
 \end{tabular}
\end{table*}

\section{Gravitational wave radiation}\label{sec:gw}

With the periodic orbits fully characterized, we now explore their observational imprints encoded in gravitational wave emission. The gravitational waveform up to quadrupole radiation can be calculated from the orbital evolution of the EMRI system using~\cite{Maselli:2021men}
\begin{equation}
	h_{ij} = \frac{4G\eta M}{c^4 D_L} \left( v_i v_j - \frac{Gm}{r} n_i n_j \right).\label{hij}
\end{equation}
Equation~\eqref{hij} involves several parameters: $M$ denotes the mass of the supermassive BH, while $m$ is the mass of the orbiting body. Given that we are dealing with an EMRI system, the condition $m \ll M$ holds. The luminosity distance of the system is $D_L$. The symmetric mass ratio is defined as $\eta = \frac{Mm}{(M+m)^2}$. Finally, the relative velocity is denoted by $\bm{v}$ and the separation vector direction by $\bm{n}$. The two tensor polarizations are constructed in a “detector-adapted” coordinate system as defined in Ref.~\cite{Poisson:2014bk}. The axes of this system are given by
\begin{equation}
 \begin{split}
	e_X &= (\cos \omega, -\sin \omega, 0), \\
	e_Y &= (\cos \iota \sin \omega, \cos\iota \cos \omega, -\sin \iota), \\
	e_Z &= (\sin \iota \sin \omega, \sin \iota \cos \omega, \cos \iota).\label{jidi}
 \end{split}
\end{equation}
Here, $\iota$ and $\omega$ stand for the inclination angle and the longitude of pericenter, respectively. Within this chosen transverse basis, the transverse-traceless tensor polarizations read~\cite{Poisson:2014bk}
\begin{equation}
 \begin{split}
 h_+ &= -\frac{2\eta}{c^4 D_L} \frac{(GM)^2}{r} (1 + \cos^2 \iota) \cos (2\varphi + 2\omega), \\
 h_\times &= -\frac{4\eta}{c^4 D_L} \frac{(GM)^2}{r} \cos \iota \sin (2\varphi + 2\omega).\label{hplus_times}
 \end{split}
\end{equation}

\begin{figure*}[htbp]
	\centering
	\includegraphics[width=0.8\textwidth,height=5cm,keepaspectratio]{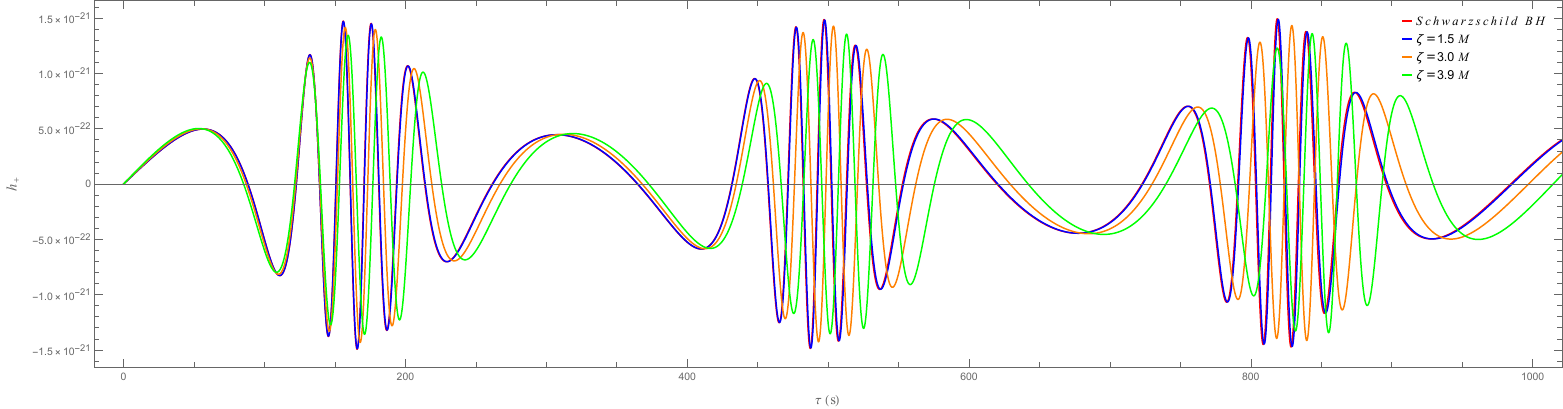}

	\vspace{0.5em}

	\includegraphics[width=0.8\textwidth,height=5cm,keepaspectratio]{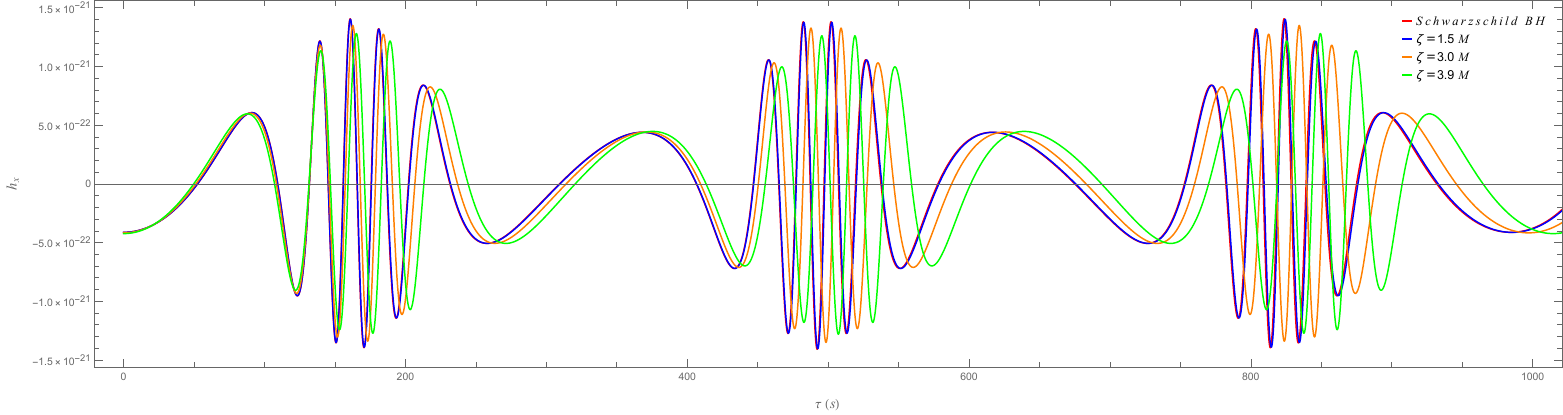}

	\caption{Waveforms from a test object of mass $m = 10M_\odot$ following the periodic orbit $(3, 1, 2)$ around a QCBH of mass $M = 10^7 M_{\odot}$, with parameters $\zeta/M = 0,\, 1.5,\, 3.0,\, 3.9$ and the specific energy fixed at $\tilde{E}=0.96$.}
	\label{Wave}
\end{figure*}

Figure~\ref{Wave} displays, as an example, the gravitational waveforms for periodic orbits with QCBH parameters $(z,w,v) = (3,1,2)$. We choose $M = 10^7 M_{\odot}$, $m=10 M_{\odot}$, $\iota=\omega=\pi/4$, and $D_{L}=200$ Mpc. Initially, the waveforms of QCBH and Schwarzschild BH almost coincide. However, after a period of evolution, the accumulated phase difference makes the waveforms distinctly different, and this difference increases with the quantum parameter $\zeta$.

\section{Properties of thin accretion disk around QCBH}\label{accretion}

The analysis of massive particle orbits and gravitational radiation provides one approach to probe the spacetime of QCBH. Accretion physics offers another complementary observational channel. In this section, we investigate its radiation characteristics by modeling thin accretion disks within the Novikov-Thorne framework~\cite{Page:1974he}. This model assumes that the accretion structure is geometrically thin, meaning the semi-thickness $H$ is negligible compared to the radial coordinate $r$ ($H \ll r$). The disk is assumed to be in a steady state with a constant mass accretion rate $\dot{M}$, maintaining hydrodynamic equilibrium where vertical pressure gradients and entropy variations are insignificant. Furthermore, we assume the disk resides in the equatorial plane and the accreting plasma moves in circular Keplerian orbits between the outer edge and the inner boundary.

\begin{figure}[htbp]
	\centering
	\includegraphics[width=8cm]{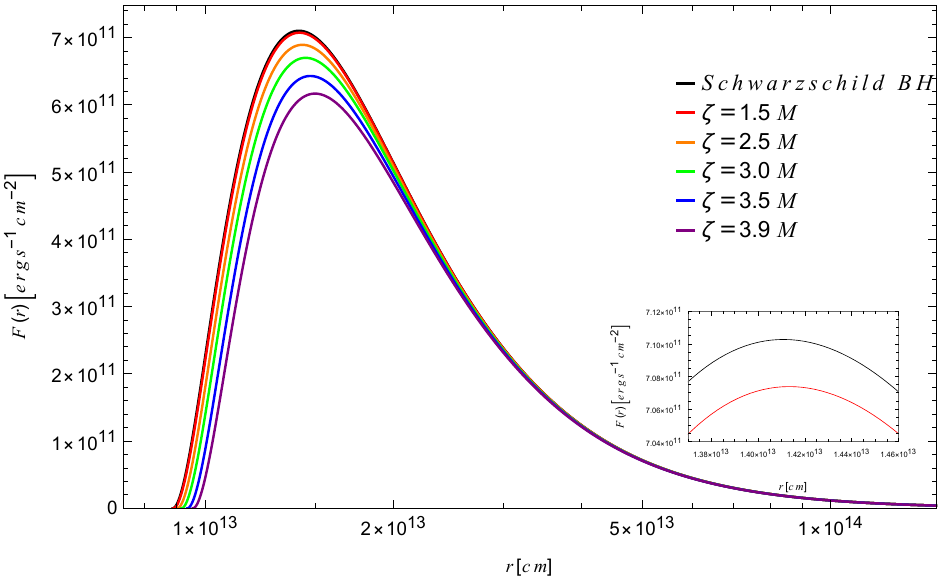}

	\caption{The energy flux $F(r)$ around the QCBH for several values of $\zeta$, where the solid black curve corresponds to the Schwarzschild BH case. The BH mass is taken as $M =10^7 M_{\odot}$, and the accretion rate is set to $\dot{M} = 2 \times 10^{-6} M_{\odot} \, \text{yr}^{-1}$.}
	\label{fig:radiation_flux}
\end{figure}

Based on the conservation laws of rest mass, energy, and angular momentum, the radiant energy flux $F(r)$ emitted from the disk surface can be derived. For a static spherically symmetric spacetime metric, the flux expression is given by~\cite{Page:1974he}:
\begin{equation}
	F(r) = -\frac{\dot{M}}{4\pi\sqrt{-g/g_{\theta \theta}}} \frac{\Omega_{,r}}{(\tilde{E}_{\rm c}-\Omega \tilde{L}_{\rm c})^2} \int_{r_{\rm ISCO}}^{r} (\tilde{E}_{\rm c}-\Omega\tilde{L}_{\rm c}) \tilde{L}_{\rm c,~r} {\rm d}r.
	\label{eq:Flux}
\end{equation}
Our study focuses on accretion driven by the QCBH with a total mass $M =10^7 M_{\odot}$ and an accretion rate $\dot{M} = 2 \times 10^{-6} M_{\odot} \, \text{yr}^{-1}$. Displayed in Fig.~\ref{fig:radiation_flux} is the energy flux $F(r)$ radiated by the accretion disk surrounding the QCBH for different values of $\zeta$. The flux distribution exhibits a characteristic pattern: it first rises, reaches a maximum, and then declines. Moreover, it can be observed that the radiative flux decreases as $\zeta$ increases. Evidently, the quantum parameter significantly influences physical observables: a larger $\zeta$ leads to a reduction in the radiation flux.

\begin{figure}[htbp]
	\centering
	\includegraphics[width=8cm]{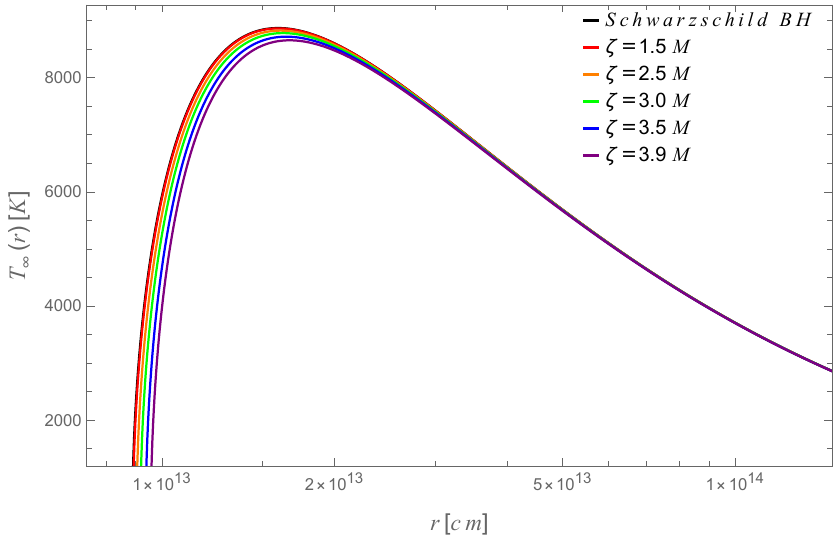}

	\caption{Curves of the thin accretion disk temperature versus the radial coordinate $r$ for different quantum parameters $\zeta$.}
	\label{fig:Temperature}
\end{figure}

Assuming the accreted matter achieves local thermodynamic equilibrium, the emission from the disk surface is modeled as blackbody radiation. According to the Stefan-Boltzmann law, the local effective temperature $T_{\rm{eff}}(r)$ is linked to the energy flux by~\cite{Page:1974he}:
\begin{equation}
 T_{\rm{eff}}(r)= \left[ \frac{F(r)}{\sigma_{\rm SB}} \right]^{1/4},
	\label{eq:Temperature}
\end{equation}
where $\sigma_{SB}$ denotes the Stefan-Boltzmann constant. For an observer at infinity, the temperature is modified by redshift effects. Incorporating the redshift factor $z$, the observed temperature $T_{\infty}(r)$ becomes:
\begin{equation}
 T_{\infty}(r)=\frac{T_{\rm{eff}}(r)}{1+z}. \label{observed_temperature}
\end{equation}
Neglecting the light bending effects, the redshift factor for a static spacetime is expressed as~\cite{Karimov:2018whx,Bhattacharyya:2000kt}:
\begin{equation}
	1+z = \frac{1 + \Omega r \sin\varphi \sin\gamma}{\sqrt{-g_{tt} - \Omega^2 g_{\varphi\varphi}}}.
	\label{eq:Redshift}
\end{equation}
Figure~\ref{fig:Temperature} shows the distribution of the radiation temperature $T_{\infty}(r)$ around the QCBH, indicating that $T_{\infty}(r)$ decreases as $\zeta$ increases.

\begin{figure}[htbp]
	\centering
	\includegraphics[width=8cm]{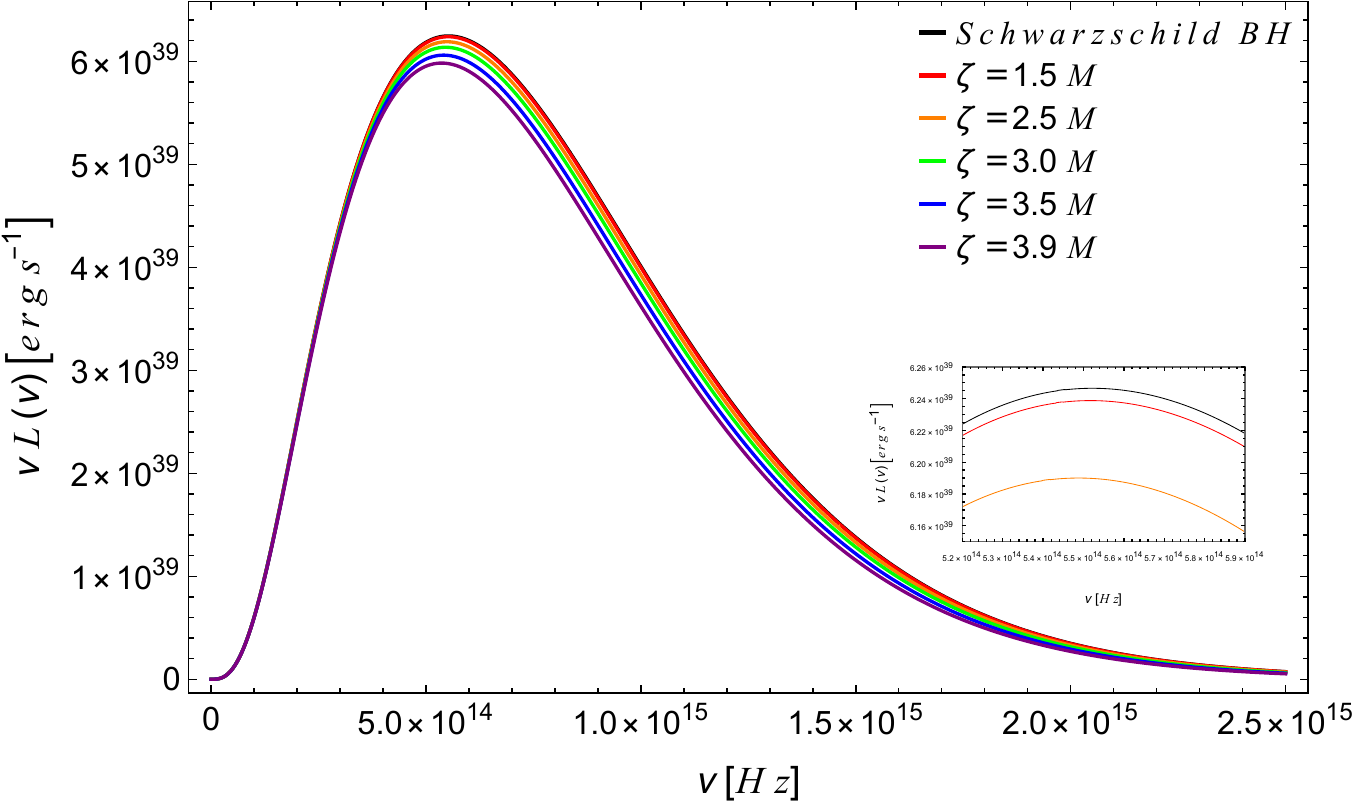}

	\caption{The emission spectrum $\nu L(\nu)$ of the accretion disk around the QCBH for different values of $\zeta$.}
	\label{fig:Spectral_Energy_Distribution}
\end{figure}

To characterize the accretion disk, we consider the spectral energy distribution (SED). The observed specific luminosity $L_{\nu,\infty}$ follows a redshifted blackbody spectrum, which is calculated by integrating the Planck function over the disk surface~\cite{Karimov:2018whx,Bhattacharyya:2000kt,Torres:2002td}:
\begin{equation}
	 L_{\nu,\infty}=4\pi d^{2}I_{\nu}=8\pi h\cos \gamma\int_{r_{\rm in}}^{r_{\rm out}}\int_{0}^{2\pi}\frac{\nu_{e}^{3}r\,\rm d r\,\rm d\varphi}{\exp\!\left(\frac{h\nu_{e}}{kT_{\rm eff}}\right)-1},
	\label{eq:Luminosity}
\end{equation}
here, $k_B$ is the Boltzmann constant, $h$ is the Planck constant, and we assume that the observer is located at a distance $d$ with an inclination angle $\gamma$. The inner boundary of the accretion disk is located at the ISCO, denoted as $r_{\rm in}=r_{\rm ISCO}$, while the outer boundary corresponds to the radius $r_{\rm out}$. The observed frequency $\nu$ is connected to the emitted frequency $\nu_e$ via the redshift factor $(1+z)$, i.e., $\nu_e =\nu(1+z)$. The spectral energy distribution at an inclination angle $\gamma = 0^\circ$, computed from Eqs.~\eqref{eq:Redshift} and \eqref{eq:Luminosity}, is shown in Fig.~\ref{fig:Spectral_Energy_Distribution}. Curves for other inclination angles are scaled proportionally by $\cos \gamma$. For frequencies $\nu \leq 5\times10^{14} \, \text{Hz}$, the spectra of the QCBH and the Schwarzschild BH are nearly identical. However, the introduction of quantum corrections suppresses the luminosity. Although the figure is plotted with $r_{\text{out}} = 100$, increasing this outer boundary radius does not produce noticeable changes.

Finally, a crucial parameter characterizing the accretion process is the radiative efficiency $\epsilon$, which quantifies the conversion capability of rest mass into radiation. Assuming that all emitted photons escape to infinity, the radiative efficiency is uniquely determined by the specific energy of the particle at the ISCO~\cite{Page:1974he}:
\begin{equation}
	\epsilon = 1 - \tilde{E}_{\rm ISCO}.
	\label{eq:Efficiency}
\end{equation}
Table~\ref{tab:Efficiency} lists the conversion efficiency $\epsilon$ for the QCBH. For the Schwarzschild BH, its theoretical radiation efficiency is $1-2\sqrt{2}/3$. For a fixed $\zeta$, the accretion process around the QCBH yields a somewhat lower $\epsilon$ than around the Schwarzschild BH. This indicates that the QCBH consistently radiates less efficiently than the standard static BH of general relativity. The lowest efficiency, $5.49\%$, is achieved by the QCBH at the maximum $\zeta$ value. As expected, when $\zeta/ M \ll 1$, the value of $\epsilon$ approaches that of the Schwarzschild BH.

\begin{table}[htbp]

	\caption{The radiation efficiency of the QCBH and the Schwarzschild BH.}
	\setlength{\heavyrulewidth}{0.08em}
	\setlength{\lightrulewidth}{0.14em}
	\setlength{\tabcolsep}{40pt}
		\begin{tabular}{cc}
			\toprule
			{\textbf{BHs}} &$\epsilon$\\
			\midrule
			{\textbf{Schwarzschild BH}} & 0.05719 \\
			QCBH ($\zeta=1.5M$) & 0.05712 \\
			QCBH ($\zeta=3.0M$) & 0.05623 \\
			QCBH ($\zeta=3.9M$) & 0.05490 \\
			\bottomrule
			\label{tab:Efficiency}
		\end{tabular}
\end{table}

\section{Conclusion}\label{conclusion}

In this paper, we performed a comprehensive analysis of the periodic orbits and astrophysical signatures associated with a QCBH without Cauchy horizons. Our investigation primarily focused on two key observational windows: the gravitational wave radiation characteristics of the EMRI system, and the electromagnetic properties of thin accretion disks.

To explore the behavior of bound orbits, we first derived the effective potential for test particles in the equatorial plane, identifying the $\tilde{E}$ and $\tilde{L}$ for circular orbits. Our analysis shows that the quantum parameter $\zeta$ significantly influences orbital stability. Notably, as $\zeta$ increases, the radii of both the ISCO and the MBO shift outward relative to the standard Schwarzschild values. The corresponding angular momentum for these critical orbits also increases with the quantum correction.

Building on the orbital framework, we further investigated periodic orbits, classifying them via the rational number $q$ and integer set $(z, w, v)$. Simulations of EMRIs around a supermassive QCBH yielded the associated gravitational waveforms. The results show that although the waveform morphology initially aligns with General Relativity predictions, a progressive phase shift accumulates over time. This phase shift scales with $\zeta$, implying that future space-based gravitational-wave detectors (e.g., LISA, Taiji, TianQin) could place constraints on the quantum correction by tracking EMRI orbital evolution.

We further evaluated electromagnetic signatures by modeling geometrically thin, optically thick accretion disks within the Novikov-Thorne framework. Radial profiles of radiant energy flux $F(r)$, temperature $T(r)$, and the emission spectrum were computed. The quantum correction consistently suppresses radiative output: both the energy flux and effective temperature of the disk decrease with increasing $\zeta$. Correspondingly, the radiative efficiency $\epsilon$ is systematically lower for the QCBH than for a Schwarzschild BH. These distinct features in the spectral energy distribution and conversion efficiency provide theoretical templates that could be tested against future high-precision observations of BH shadows and accretion flows.

\begin{acknowledgements}
	This work is supported in part by NSFC Grants No. 12165005 and No. 11961131013.
\end{acknowledgements}

\noindent{\bf Data Availability Statement} This manuscript has no associated data. [Authors' comment: Data sharing not applicable to this article as no datasets were generated or analyzed during the current study.]

\vspace{0.2cm}
\noindent{\bf Code Availability Statement} This manuscript has no associated code/software. [Authors' comment: Code/Software sharing not applicable to this article as no code/software was generated or analyzed during the current study.]



\end{document}